\documentclass[journal]{IEEEtran}
\usepackage{hyperref}
\usepackage{amsmath, amssymb}
\usepackage{graphicx}
\usepackage[export]{adjustbox}
\usepackage{xcolor}
\usepackage{caption}
\usepackage{subcaption}
\usepackage{acronym}
\usepackage{array}
\usepackage{multirow}
\usepackage{enumitem}
\usepackage{gensymb}

\graphicspath{{images/}}

\def\loss{{\mathcal L}}

\def\C{{\bf C}}
\def\c{{\bf c}}
\def\f{{\bf f}}
\def\X{{\bf X}}
\def\x{{\bf x}}
\def\y{{\bf y}}
\def\Y{{\bf Y}}
\def\T{{\bf T}}
\def\t{{\bf t}}

\def\bphi{{\boldsymbol \phi}}
\def\btheta{{\boldsymbol \theta}}

\def\D{{\mathcal{D}}}

\def\Dcc{{\mathcal{D}_{\mathrm{c} \mathrm{\hat c}}}}
\def\Dx{{\mathcal{D}_{\mathrm{x}}}}
\def\Dxx{{\mathcal{D}_{\mathrm{x} \mathrm{\hat x}}}}
\def\Dtt{{\mathcal{D}_{\mathrm{t} \mathrm{\hat t}}}}
\def\Dt{{\mathcal{D}_{\mathrm{t}}}}

\def\pt_a{p({\bf t}|{\bf a})}

\def\pdx{p_{\mathcal D}({\bf x})}

\newcommand{\argmin}{\operatornamewithlimits{argmin }}
\newcommand{\argmax}{\operatornamewithlimits{argmax }}

\newcolumntype{P}[1]{>{\centering\arraybackslash}p{#1}}

\ifCLASSINFOpdf
\else
\fi
\begin{document}
%
\title{Mobile authentication of copy detection patterns}
%
%
%

\author{Olga~Taran, Joakim~Tutt, Taras~Holotyak,~\IEEEmembership{Member,~IEEE,} Roman~Chaban, Slavi~Bonev, Slava~Voloshynovskiy,~\IEEEmembership{Senior Member,~IEEE}
\thanks{S. Voloshynovskiy is a corresponding author.} 
\thanks{This research was partially funded by the Swiss National Science Foundation SNF No. 200021\_182063.}\\
\IEEEauthorblockA{Department of Computer Science, University of Geneva, Switzerland \\
\{olga.taran, joakim.tutt, taras.holotyak, roman.chaban, slavi.bonev, svolos\}@unige.ch}
}

\maketitle

\begin{abstract}
In the recent years, the copy detection patterns (CDP) attracted a lot of attention as a link between the physical and digital worlds, which is of great interest for the internet of things and brand protection applications. However, the security of CDP in terms of their reproducibility by unauthorized parties or clonability remains largely unexplored. In this respect this paper addresses a problem of anti-counterfeiting of physical objects and aims at investigating the authentication aspects and the resistances to illegal copying of the modern CDP from machine learning perspectives. A special attention is paid to a reliable authentication under the real life verification conditions when the codes are printed on an industrial printer and enrolled via modern mobile phones under regular light conditions. The theoretical and empirical investigation of authentication  aspects of CDP is performed with respect to four  types of  copy  fakes from the point of view of (i) multi-class supervised classification as a baseline approach and (ii) one-class  classification as a real-life application case.  The  obtained  results  show  that  the modern machine-learning approaches and the technical capacities of modern mobile phones allow to reliably authenticate CDP on end-user mobile phones under the considered classes of fakes\footnote{The partial results from this work were submitted to the special session on "Forensics and Security of Physical Objects" of the IEEE International Workshop on Information Forensics and Security 2021.}.

\end{abstract}


%
\IEEEpeerreviewmaketitle


\section{Introduction}
%
%
%
%
\IEEEPARstart{I}{n} the modern world of globally distributed economy it is extremely challenging to ensure a proper production, shipment, trade distribution, consumption and recycling of various products and goods of physical world. These products and goods range from everyday food to some luxury objects and art. Creation of digital twins of these objects with appropriate track and trace infrastructures complemented by cryptographic tools like blockchain represents an attractive option. However, it is very important to provide a robust, secure and unclonable link between a physical object and its digital representation in centralized or distributed databases. This link might be implemented via overt channels, like personalized codes reproduced on products either directly or in a form of coded symbologies like 1D and 2D codes or covert channels, like invisible digital watermarks embedded in images or text or printed by special invisible inks. However, many codes of this group are easily copied or can be regenerated. Thus, there is a great need in unclonable modalities that can be easily integrated with the printable codes. This necessity triggered the appearance and growing popularity of Printable Graphical Codes (PGC). During the last decade, the PGC attracted many industrial players and governmental organizations. One of the most popular nowadays type of PGC is a union of traditional 2D codes and \textit{copy detection patterns} (CDP) \cite{picard2004digital}. 

\begin{figure}[t!]
	\centering
    \includegraphics[width=0.95\linewidth]{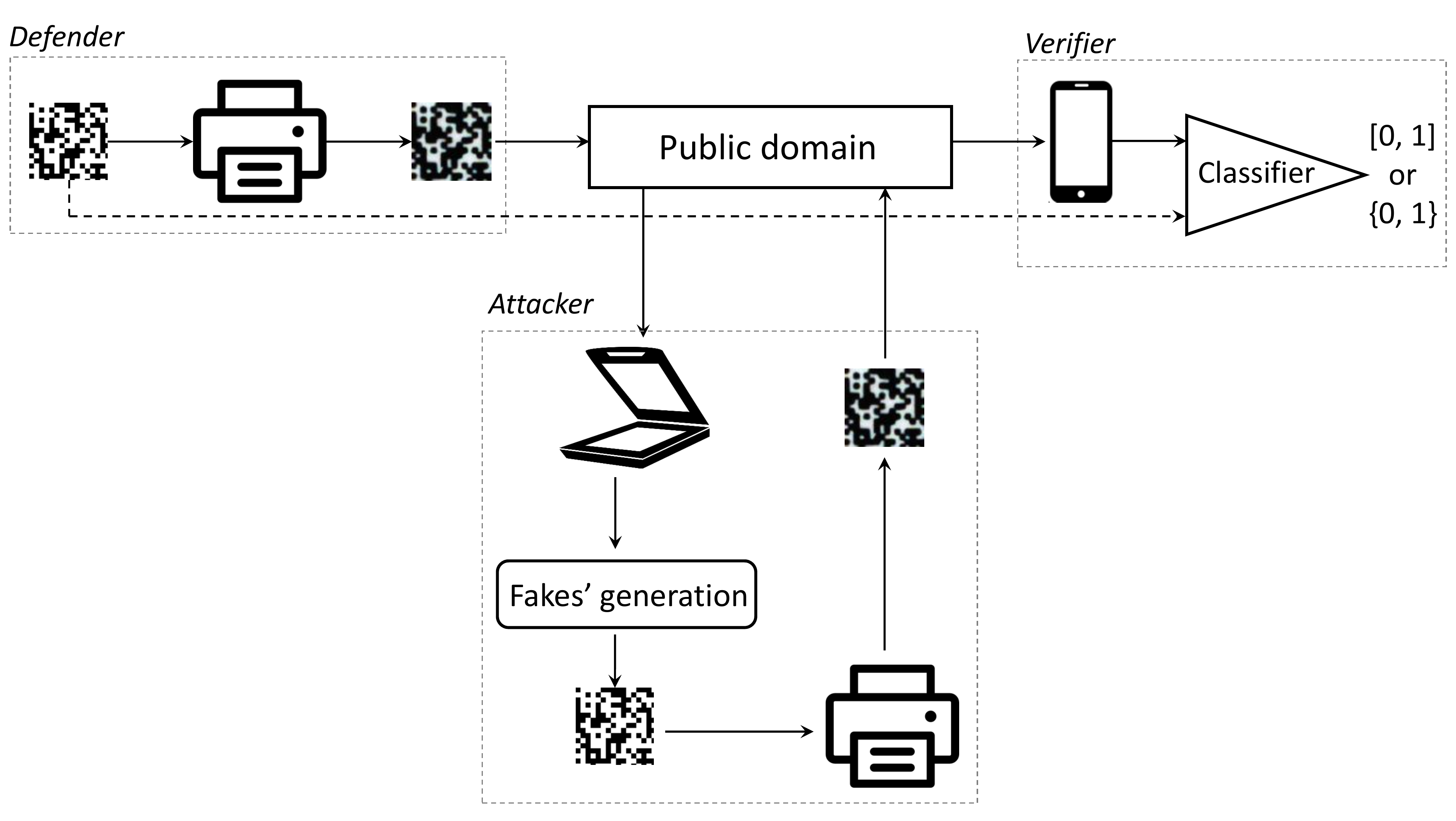}
    \caption{General scheme of the CDP life cycle starts from the generation of the digital templates by the defender and their following printing. The produced codes go to the public domain. An attacker has an access to the publicly available printed codes and can produce different type of fakes that are then also distributed in the public domain. A verifier should digitize the printed codes from the public domain and validate them via some classifier. As it is shown by the dashed line, the validation might be produced with or without taking the digital templates into account. For the defender-verifier pair the main goal is to minimize the probability of error. In contrast, the attacker aims at maximizing the probability of error.}
    \label{fig:pgc life cycle}
\end{figure}

General scheme of the CDP life cycle is shown in Fig. \ref{fig:pgc life cycle}. The CDP security is based on a so-called information loss principle: each time the code is printed or scanned some information about the original digital template is inevitably lost. 
In the case of printable codes, the information loss principle is based on physical phenomena of random interaction between the ink or toner with a substrate. As a result any dot undergoes a complex unpredictable modification and changes its shape accordingly to a dot gain effect. Generally, the black dot increases in its size. A white hole on a black background accordingly decreases its area due to the dot gain of nearest black dot surround. 

In the case of image acquisition the information loss principle refers to a loss of image quality due to various factors that include variability of illumination, finite and discrete nature of sampling in CCD/CMOS sensors, non-linearity  in sensor sensitivity, sensor noise and various sensor defects, etc. All together, the enrolled image is characterized by some variability that degrades the quality of image in terms of its correspondence to the original digital template from which the code was printed.

Nowadays, there exists a big variety of different approaches aiming to combine CDP and widely used traditional 2D codes. Without pretending to be exhaustive in the presented overview, some of the most representative approaches are mentioned below.

\begin{figure}[t!]
	\centering
	
	\begin{subfigure}{0.3\columnwidth}
		\centering
        \includegraphics[width=0.8\linewidth,valign=t]{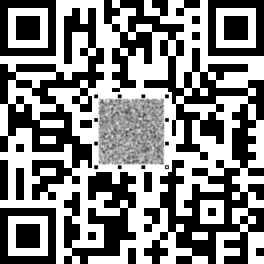}
	    \caption{\centering One-level PGC \cite{picard2021counterfeit}.}
	    \label{fig:one level pgc}
	\end{subfigure}		
	\begin{subfigure}{0.32\columnwidth}
		\centering
        \includegraphics[width=0.75\linewidth,valign=t]{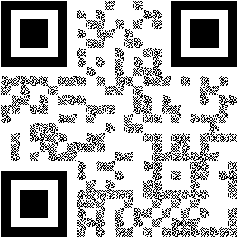}
		\caption{\centering Two level 2LQR codes \cite{tkachenko2016printed}.}
		\label{fig:pgc 2lqr}
	\end{subfigure}	
	\hspace{0.2cm}	
	\begin{subfigure}{0.32\columnwidth}
		\centering
        \includegraphics[width=1\linewidth,valign=t]{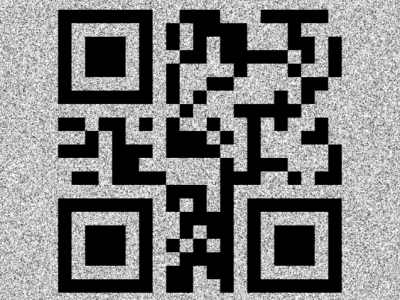}
		\caption{\centering Two level W-QR codes \cite{nguyen2017watermarking}.}
		\label{fig:pgc wqr}
	\end{subfigure}	
    \caption{Examples of different types of modern PGC with CDP modulations.}
    \label{fig:pgc examples}        
\end{figure}

In general, it is possible to distinguish the standard one-level PGC and more advanced multi-level PGC. Examples of these codes are given in Fig. \ref{fig:pgc examples}. The one-level PGC is shown in Fig. \ref{fig:one level pgc}. According to the presented design, a CDP central part is inserted into a structure of 2D QR-code \cite{picard2021counterfeit}. Originally the multi-level PGC aimed at increasing the storage capacity of the regular PGC \cite{villn2006multilevel}. Recently, the multi-level PGC are considered as a tool to increase the security of standard PGC. Without loss of generality, it is possible to identify the multi-level PGC with a modulation of the main black symbols as shown in Fig. \ref{fig:pgc 2lqr} and a background modulation as illustrated in Fig. \ref{fig:pgc wqr}. 

The most well known multi-level PGC of the first type are so-called two level QR (2LQR) codes proposed in \cite{tkachenko2015two,tkachenko2016printed}, where the standard black modules are substituted by special modulated patterns. The general principles of modulation of multi-level codes were initially considered and theoretically analysed in \cite{Villan:TIFS2006}. The public level of this code is read as normal standard QR code. The texture patterns are chosen to be sensitive to the print \& scan process. At the same time, the modulation pattern can carry out private message. Furthermore, the idea of 2LQR was extended in \cite{cheng2018new} by the use of different encrypting strategies. 
The anti-counterfeiting performance of these codes was mainly tested based on desktop printers and scanners \cite{tkachenko2015two,tkachenko2016printed}. Thus, there is a great interest in validation of these codes under the industrial printing and mobile phone authentication.

The second type of multi-level PGC is so-called W-QR codes proposed in \cite{nguyen2017watermarking}, where the authors substitute the background of a standard QR code by a specific random texture. The embedded texture does not affect the readability of the standard code but it should be sensitive to the print \& scan process in such a way to give a possibility to authenticate the original code from the counterpart. The authors propose a particular random textured pattern, which has a stable statistical behavior.  
Thus, the attacker targets to estimate the parameters of the used textured pattern.


Despite the differences in ways how the traditional QR codes and CDP are combined, in general case, the authentication of digital artwork based on the CDP is done by comparing  the reference template with the printed version scanned using a scanner or camera of mobile phone. As a reference template there can be used either a digital template or enrolled printed version of the same artwork. The comparison can  be done in different ways either in the spatial or frequency domain using a correlation, distance metrics or a combined score of different features, etc., \cite{picard2004digital, dirik2012copy}. Alternatively, one can also envision an authentication in a transform domain using latent space of pretrained classifiers or auto-encoders \cite{Taran2020icassp}. 

Despite a great interest, the robustness of CDP, used in PGC, to the copy attacks remains a little studied problem. Therefore, the current work is dedicated to the investigation of the authentication aspects of CDP under industrial settings from the perspective of modern machine learning. 

The main contributions of this paper are:
%

\begin{itemize}
    \item We provide the extended representation of production and enrollment procedures and settings of the Indigo mobile dataset of CDP created under the regular industrial settings and briefly presented in \cite{Taran2021IndigoMobile}.
    \item We provide an extention of the multi-class supervised classification results presented in \cite{Taran2021IndigoMobile}. Namely, in addition to the supervised classifier trained in the binary (or two classes) setup with respect to the different types of the fakes, we provide new results of the performance of supervised classifier trained in three and five classes classification setups.
    \item We investigated the authentication aspects of the CDP from the perspective of one-class classification in the spatial domain with respect to the different type of reference codes: the digital templates and the physical references.
    \item For the one-class classification in the deep processing domain we provide more detailed mathematical explanation of the model under investigation.
    \item In addition to the five basic scenarios of the one-class classification based on the one-class SVM, we provide more deep investigation of the problem under investigation with respect to the Hamming distance descision criteria. Also, we provide more detailed analysis of the latent space of the deep models under investigation.
	\item Finally, we investigate the complexity of the main models under investigation.
\end{itemize}

\noindent\textbf{Notations:} We use the following notations: $\t \in \{0, 1\}^{m \times m}$ denotes an original digital template; $\x \in \mathbb{R}^{m \times m}$ corresponds to an original printed code, while $\f \in \mathbb{R}^{m \times m}$ is used to denote a printed fake code; $\y \in \mathbb{R}^{m \times m}$ stands for a probe that might be either original or fake. $p_t(\t)$ and $p_\D(\x)$ correspond to empirical data distributions of the digital templates and original printed codes, respectively. The discriminators corresponding to Kullback–Leibler divergences are denoted as $\Dx$, where the subscript indicates the space to which this discriminator is applied to.  

\section{Datasets}
\label{chap4_sec:pgd datasets}

\subsection{State-of-the-art datasets}

The majority of the research experiments in the domain of CDP are performed either on synthetic data or on small private datasets. The production of datasets of real CDP is a very time consuming  and  quite  costly process. It requires the printing and acquisition of the original CDP, the production and acquisition of fakes preferably on the equipment close to the industrial one. 

\begin{figure}[t!]
	\centering
	\begin{subfigure}{0.49\columnwidth}     
		\centering	
        \includegraphics[width=0.5\textwidth,valign=t]{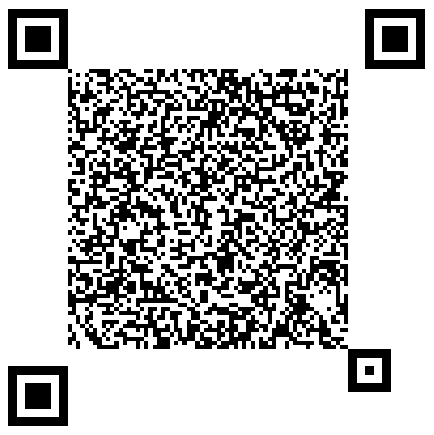}
        \caption{Binary digital template.}
        \label{fig:indigo mobile ex digital template}
	\end{subfigure}
	\hfill
	\begin{subfigure}{0.49\columnwidth}
		\centering
        \includegraphics[width=0.5\textwidth,valign=t]{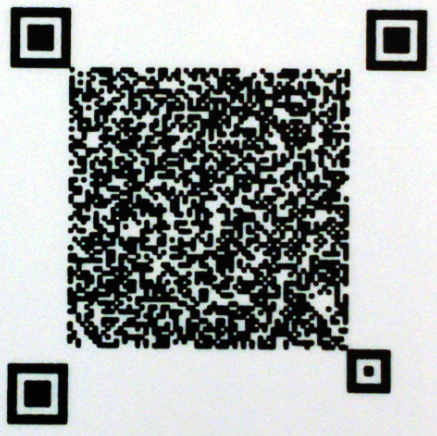}
        \caption{Printed original code.}
        \label{fig:indigo mobile ex original}
	\end{subfigure}
    \caption{Examples of (a) a binary digital template used for printing and (b)  the printed original code from the Indigo mobile dataset enrolled by the mobile phone.}
    \label{fig:indigo_mobile_photo_example}        
\end{figure}


Up to our best knowledge, there are only few publicly available datasets that were created to investigate the clonability aspects of CDP: 
\begin{itemize}
	\item[(1)] The DP0E \cite{Taran2019icassp} and its extension DP1E \& DP1C \cite{Taran2020icassp} are the datasets of real and counterfeited CDP based on \textit{DataMatrix} modulation \cite{ISO2006} printed at resolution 1200 dpi with four  printers: two laser (a) Samsung Xpress 430 and (b) Lexmark CS310 and two Inkjet (c) Canon PIXMA  iP7200 and (d) HP OfficeJet Pro 8210. The enrollmen was performed by using the high resolution scanners at resolution 1200 ppi: Canon 9000F and Epson V850 Pro. The DP1E \& DP1C dataset contains 6528 codes produced from 384 digital templates with symbol size $6 \times 6$ elements, with 3072 printed original codes and 3072 fake codes printed on the same printers as original codes.
	\item[(2)] The CSGC dataset \cite{yadav2019estimation} consists of 3800 codes produced from 950 digital templates with symbol size $1 \times 1$ elements and 2850 original codes printed on the Xerox Phaser 6500 laser at resolution 600 dpi and scanned by the Epson V850 Pro scanner under three resolutions: 2400 ppi, 4800 ppi and 9600 ppi.
	\item[(3)] Indigo mobile dataset \cite{Taran2021IndigoMobile} contains the CDP printed on the industrial printer HP Indigo 5500 DS at resolution 812 dpi. This dataset was created to investigate the authentication capabilities of CDP under conditions closer to the real life environment. In this respect, instead of high quality scanners the printed codes were enrolled by a mobile phone \textit{iPhone XS} under regular room light conditions. The dataset contains 300 digital templates with symbol size $5 \times 5$ elements, 300 printed original codes and 1200 typical copy-fake codes.

\end{itemize}

As an example of the real-life scenario the Indigo mobile dataset presents a particular interest for the detailed practical investigation.

\begin{figure}[t!]
    \centering
    \includegraphics[scale=0.5]{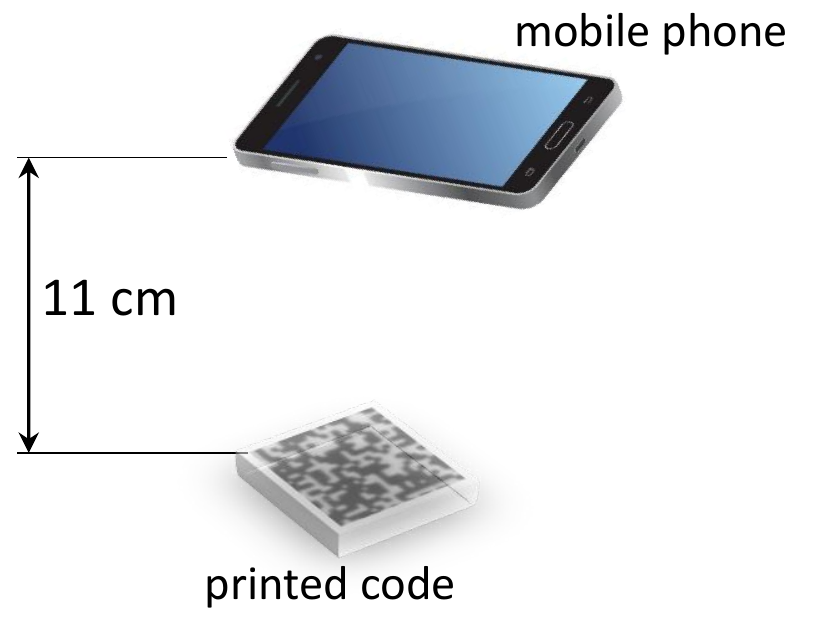}
    \caption{The schematic representation of the mobile phone acquisition setup.}
    \label{fig:mobile_acquisition}
\end{figure}
\subsection{Indigo mobile dataset}
\label{ch4_subsec:auth dataset}

\begin{figure}[t!]
	\centering
	\begin{subfigure}{0.32\columnwidth}     
		\centering	
        \includegraphics[width=0.7\linewidth,valign=t]{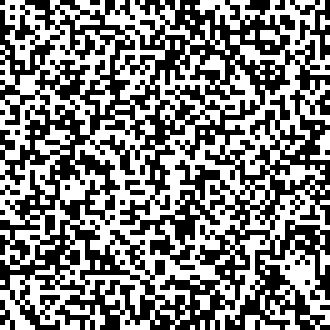}
        \caption{Digital template.}
        \label{fig:digital template}
	\end{subfigure}
	\begin{subfigure}{0.32\columnwidth}
		\centering
        \includegraphics[width=0.7\linewidth,valign=t]{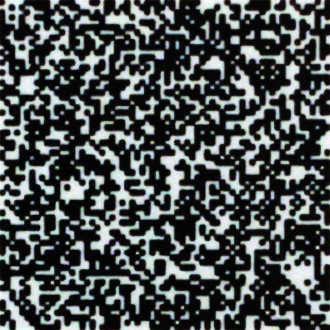}
        \caption{Original.}
        \label{fig:original}
	\end{subfigure}
	\begin{subfigure}{0.32\columnwidth}
		\centering
        \includegraphics[width=0.7\linewidth,valign=t]{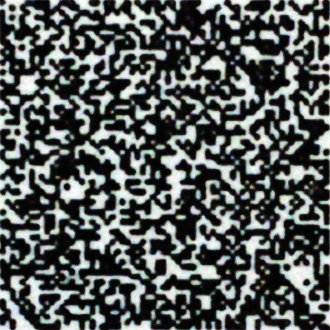}
        \caption{Fake \#1 white.}
        \label{fig:f1w}
	\end{subfigure}	
	\\
	\begin{subfigure}{0.32\columnwidth}
		\centering
        \includegraphics[width=0.7\linewidth,valign=t]{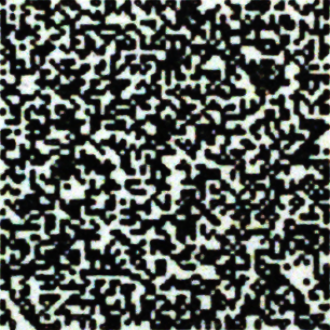}
        \caption{Fake \#1 gray.}
        \label{fig:f1g}
	\end{subfigure}	
	\begin{subfigure}{0.32\columnwidth}
		\centering
        \includegraphics[width=0.7\linewidth,valign=t]{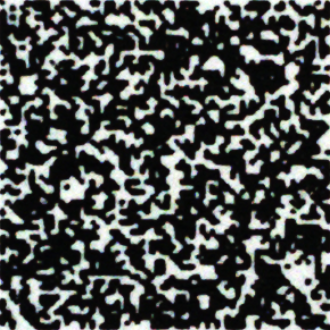}
        \caption{Fake \#2 white.}
        \label{fig:f2w}	
    \end{subfigure}    
	\begin{subfigure}{0.32\columnwidth}
		\centering
        \includegraphics[width=0.7\linewidth,valign=t]{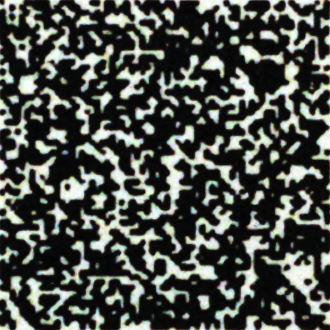}
        \caption{Fake \#2 gray.}
        \label{fig:f2w}
	\end{subfigure}	
    \caption{Examples of original and fake codes with symbol size $5 \times 5$ elements taken by a mobile phone from the Indigo mobile dataset. }
    \label{fig:auth_db codes visualisation}        
\end{figure}

Indigo mobile dataset includes 300 distinct digital \textit{DataMatrix} templates $\t \in \{0, 1\}^{330 \times 330}$ with the symbols of size $5 \times 5$ elements\footnote{To ensure accurate symbol representation, each printed symbol should be represented by at least $3 \times 3$ pixels. Taking into account the difference between the industrial printing resolution (about 812 dpi) and the average resolution of the mobile phones (about 600 - 900 ppi) especially in the development countries, where the problem of counterfeiting is particularly important, on can estimate the symbol size from about $4 \times 4$ till $5 \times 5$ pixels.}. An example of the digital template is given in Fig. \ref{fig:indigo mobile ex digital template}. The digital templates consist of the central CDP and four synchro-markers that allow to make an accurate synchronization and cropping of the code of interest. To simulate the real life scenario, the generated digital templates were printed on the industrial printer \textit{HP Indigo 5500 DS} at the resolution 812 dpi\footnote{It should be pointed out that the native printing resolution of HP Indigo 5500 DS is 812.8 dpi. The impact of printing resolution and the symbol size is a subject of our ongoing research.}. The acquisition of the printed codes is performed under regular room light using mobile phone \textit{iPhone XS} (12 Mpixels) under the automatic photo shooting settings in Lightroom application\footnote{\url{https://apps.apple.com/us/app/adobe-lightroom-photo-editor/id878783582}}. The mobile phone is held parallel to the printed code at height 11 cm as schematically shown in Fig \ref{fig:mobile_acquisition}. The photos are taken in DNG format to avoid built-in mobile phone image post-processing. An example of obtained photo is shown in Fig. \ref{fig:indigo mobile ex original}. The following cropping of the code is performed in an automatic way by applying a geometrical synchronization with four squared synchro-markers. Finally, the cropped codes are converted to the RGB format\footnote{\url{https://docs.opencv.org/4.5.2/d8/d01/group__imgproc__color__conversions.html\#ga397ae87e1288a81d2363b61574eb8cab}}. The obtained codes are $\x \in \mathbb{R}^{330 \times 330}$ with symbols' size $5 \times 5$ elements. Examples of the obtained code is shown in Fig. \ref{fig:original}. 

To simulate typical scenario for an unexperienced counterfeiter, we produce copies based on standard copy machines. The two different copy machines in copy regime "text" are used: \textit{(1)} RICOH MP C307 and \textit{(2)} Samsung CLX-6220FX. The fakes are produced on two types of paper: white paper 80 g/m\textsuperscript{2} and gray paper 80 g/m\textsuperscript{2}.

Thus, as it is mentioned in \cite{Taran2021IndigoMobile}, the four fake codes for each original printed code were produced, namely: 
\begin{enumerate}[noitemsep] 
	\item \textit{Fakes \#1 white}: made by the copy machine (1) on the white paper.
	\item \textit{Fakes \#1 gray}: made by the copy machine (1) on the gray paper.
	\item \textit{Fakes \#2 white}: made by the copy machine (2) on the white paper.
	\item \textit{Fakes \#2 gray}: made by the copy machine (2) on the gray paper.
\end{enumerate}

To be coherent with the enrolled original printed codes, the acquisition of the produced fakes is performed in the same way using the same mobile phone under the same photo and light settings as for the  original printed codes. 

In total, the Indigo mobile dataset contains 1800 codes: 300 distinct digital templates; 300 enrolled original printed codes and 1200 enrolled fake printed codes: 300 originals $\times$ 4 type of fakes\footnote{The Indigo mobile dataset will be publicly available upon the paper acceptance. The web link will be provided here.}.

Examples of the obtained digital, original and fake codes are shown in Fig. \ref{fig:auth_db codes visualisation}. Due to a built-in morphological processing of the Ricoh copy machine the fakes \#1 are more accurate with a dot gain close to the original codes. In the case of the fakes \#2 the dot gain is much higher and, as a result, the symbols contain more black ink and look darker. Visually the difference between the two types of used paper is not evident.


For the empirical evaluation the Indigo mobile dataset was split into three sub-sets: \textit{training} with 40\% of data, \textit{validation} with 10\% of data and 50\% of data is used for the \textit{test}. To avoid the bias in the choice of training and test data, each investigated model was trained five times under randomly splitting data between these subsets. Moreover, the following data augmentations were used: \textit{(i)} the rotations on 90\degree, 180\degree and 270\degree; \textit{(ii)} the gamma correction with variable function $(.)^\gamma$, where $\gamma \in [0.5, 1.2]$ with step $0.1$ is the parameter of gamma correction.

\section{Multi-class supervised classification}
\label{sec_chap4: supervised classification}

\begin{figure}[t!]
	\centering
    \includegraphics[width=0.7\linewidth]{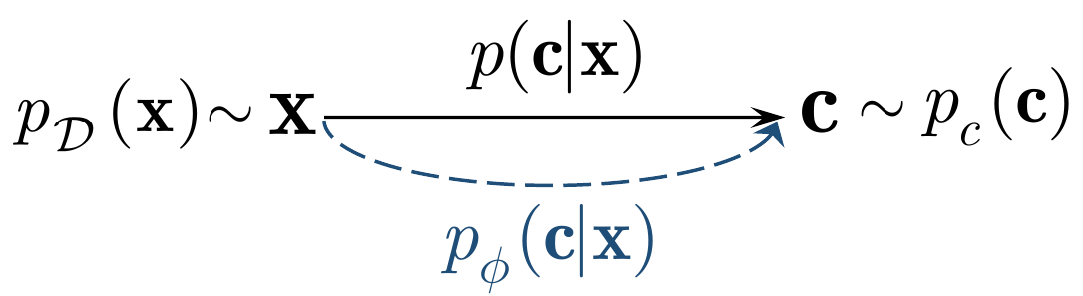}
    \caption{Supervised classification problem: general formulation.}
    \label{fig:classification ib}
\end{figure}

\begin{table*}[t!]
\begin{minipage}{\textwidth}
	\centering
	\renewcommand*{\arraystretch}{1.25}
	\caption{The classification error of the supervised multi-class classifier (in \%).}	\label{tab:supervised five class classification}
	{\small	
	\begin{tabular}{p{0.14\textwidth}|P{0.14\textwidth}|P{0.14\textwidth}|P{0.14\textwidth}|P{0.14\textwidth}|P{0.14\textwidth}} \hline
	Classification type & Originals & Fakes \#1 white & Fakes \#1 gray & Fakes \#2 white & Fakes \# 2 gray \\ \hline
	2-class\protect\footnote{$P_e$ corresponds to the $P_{miss}$ for the originals and to the $P_{fa}$ for the fakes.} & 0.00 & \multicolumn{4}{c}{0.28 ($\pm$0.3)} \\ \hline
	3-class & 0.00 & \multicolumn{2}{c}{0.78 ($\pm$0.68)} & \multicolumn{2}{|c}{0.35 ($\pm$0.5)} \\	\hline	
	5-class & 0.00 & 23.26 ($\pm$7.55) & 21.56 ($\pm$0.81) & 16.88 ($\pm$6.62) & 11.35 ($\pm$4.89) \\ \hline	
	\end{tabular}
	}	
\end{minipage}	
\end{table*}


\subsection{Theoretical analysis}

The supervised multi-class classification is chosen as a base-line to validate the authentication efficiency of CDP. The complete availability of fakes at the training stage for the classification gives the defender an information advantage over the attacker. Such a scenario is an ideal case for the defender and the worst case for the attacker. It assumes that, besides the original digital templates $\{\t_i\}_{i=1}^M$ and the corresponding printed codes $\{\x_i\}_{i=1}^M$, the defender has an access to the fake codes $\{\f_i\}_{i=1}^{M_f}$, $M_f \le M$. 

From the information-theoretic point of view the problem of a supervised classifier training given the labeled data $\{\x_i, \c_i\}^N_{i=1}$ generated from a joint distribution $p(\x, \c)$  is formulated as a training of a parameterized network $p_\bphi(\c|\x)$ that is an approximation of $p(\c|\x)$ originating from the chain rule decomposition $p(\x, \c) = p_\D(\x) p(\c|\x)$. The training of the network $p_\bphi(\c|\x)$ is performed based on the maximisation of a mutual information $I_{\bphi}(\X;\C)$ between $\x$ and $\c$ via $p_\bphi(\c|\x)$:
\begin{equation} 
    \hat{\bphi} = \argmax_{\bphi} I_{\bphi}(\X;\C),
    \label{eq:ib supervised classification max}     
\end{equation}
%
that can be rewritten as: 
%
\begin{equation} 
    \hat{\bphi} = \argmin_{\bphi}  \loss_{\textrm{Supervised}}(\bphi ),
    \label{eq:ib supervised classification min}     
\end{equation}
%
where $\loss_{\textrm{Supervised}}(\bphi ) =  - I_{\bphi}(\X;\C)$.

The mutual information in (\ref{eq:ib supervised classification max}) is defined as: 
%
\begin{equation}
	\begin{aligned} 
		I_\bphi(\X; \C) & \triangleq 
		%
		\mathbb{E}_{p(\x,\c)} \left[ \log \frac{p_\bphi(\c|\x)}{p_c(\c)}  \right] \\
		%
		& =  \underbrace{\mathbb{E}_{p(\x,\c)} \left[ \log p_\bphi(\c|\x) \right] }_\text{$\Dcc$} - \underbrace{\mathbb{E}_{p_c(\c)} \left[ \log p_c(\c)  \right]}_\text{= constant},
	 \end{aligned} 
	\label{eq:ib-occ first term direct decomposition}
\end{equation}
%
where $H(\C) = -\mathbb{E}_{p_c(\c)} \left[ \log p_c(c)  \right]$ is the entropy of $\c$ and it is a constant that does not depend on $\bphi$. 

Therefore, the optimisation problem (\ref{eq:ib supervised classification min}) reduces to: 

\begin{equation} 
\begin{aligned}
\hat{\bphi} & = \argmin_{\bphi}  \loss_{\textrm{Supervised}}(\bphi) = \argmin_{\bphi} -\Dcc.
\end{aligned}
\label{eq:ib supervised classification min}     
\end{equation}

\textbf{Remark:} In practice the $\Dcc$ term is optimized with respect to the cross-entropy loss.

\subsection{Experimental results}

The performance of the presented model (\ref{eq:ib supervised classification min}) was empirically evaluated on the Indigo mobile dataset\footnote{To ensure reproducible research the python code for all investigated models with the description of used training parameters will be publicly available upon the paper acceptance. The github link will be provided here.}. The supervised multi-class classification is performed in two scenarios: \textit{(1)} multi-class classification and \textit{(2)} binary classification.

\subsubsection{Multi-class classification}
\label{subsec_chap4:five class classification}

The multi-class supervised classification aims at investigating the performance of the base-line supervised classification scenario, where the model is trained on all classes of the data. Therefore, it corresponds to the case of the informed defender who knows all types of fakes in advance. At the inference stage, three validation scenarios are evaluated:

\begin{itemize}
	\item 5-class classification: the ability of the model to distinguish all classes of the data, i.e., originals and four types of fakes.	
	\item 3-class classification: the ability of the model to distinguish the originals, fakes from the first (fakes \#1) and the second (fakes \#2) groups.	
	\item 2-class classification: the ability of the model to distinguish the originals from all types of fakes considered as a joint class.
\end{itemize}

Due to the relatively small amount of the codes in the Indigo mobile dataset and to avoid the bias in the selection of data for training and testing, the classification model is trained five times on the randomly chosen subset of data.

At the inference stage, the query sample $\y$, which might be either the original code $\x$ or one of the fakes $\f^k$, $k = 1, ..., 4$, is passed through a deterministic classifier $g_\bphi$ such that $p_\bphi(\c|\x) = \delta(\c - g_\bphi(\x))$ and $\delta(.)$ denotes the Dirac delta-function or simply $\c = g_\bphi(\x)$. Each class is encoded as one-hot-encoding with the class $i^{\textrm{th}}$ represented as $\c_i = [0, ..., 1, ..., 0]^\textrm{T}$, with "1" in the position of $i^{\textrm{th}}$. Herewith, $g_\bphi$ is trained with respect to the term $\Dcc$ in (\ref{eq:ib supervised classification min}). The term $\Dcc$ represents the cross-entropy in this case. The obtained classification error $P_{e} = Pr[\hat{\c} \neq \bf{C} | \bf{C} = \c]$ is given in Table \ref{tab:supervised five class classification}. It is easy to see that the investigated model is capable to authenticate the original codes without mistakes in all considered scenarios. 

\begin{figure}[t!]
	\centering
    \includegraphics[width=0.65\linewidth]{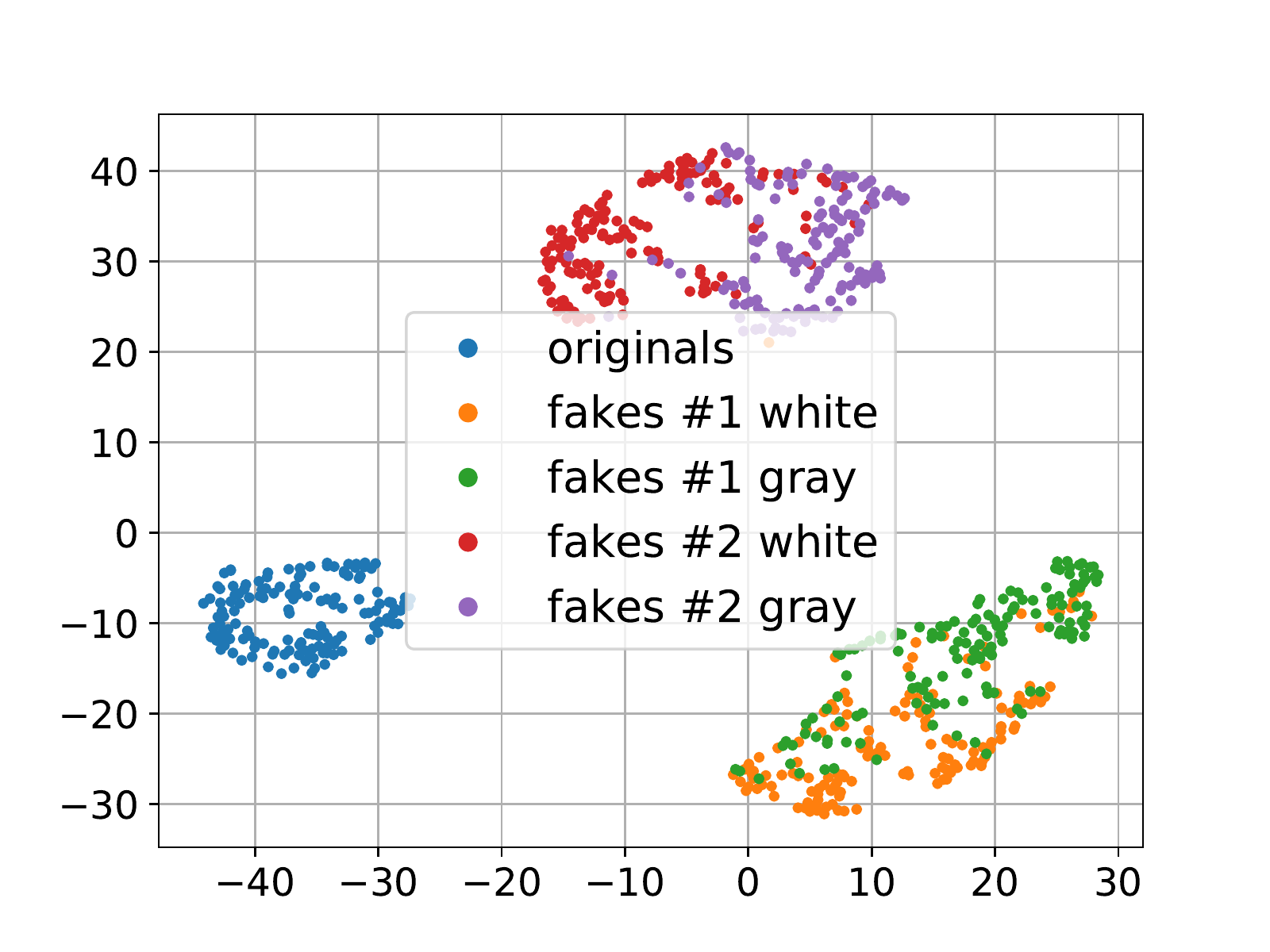}    
    \caption{T-SNE of the latent space (the last layer before an activation function) of the supervised classifier trained on originals and all type of fakes.}
    \label{fig:tsne five classes classification}
\end{figure}

The classification error about $0.28\%$ in the two classes validation setup ("2-class" label in Table \ref{tab:supervised five class classification}) indicates that despite the visual similarity the classifier is capable to distinguish original and fakes with high enough accuracy. From the three classes validation scenario ("3-class" label in Table \ref{tab:supervised five class classification}), one can notice that the model confuses more the fakes \#1 than fakes \#2. The last validation scenario ("5-class" label in Table \ref{tab:supervised five class classification}) shows that for both groups of fakes the most difficult is to distinguish between the white and gray paper type of fakes. In addition, in Fig. \ref{fig:tsne five classes classification} the t-SNE visualization \cite{hinton2002stochastic} of the latent space (the last layer before an activation function) of the classifier trained in 5-class classification scenario is illustrated. From that visualization one can easily see the same phenomena: three main classes (originals, fakes \#1 and fakes \#2) are well separated while the samples printed on the white and gray papers overlap. This indicates that the substrate identification is a difficult problem even for the supervised classifier under the considered imaging setup.

\subsubsection{Binary classification}
\label{subsec_chap4:two class classification}

The supervised binary classification aims at investigating the influence of the fakes' type used for the training on the model efficiency at the inference stage. In this respect, the training is performed separately on each type of fakes. Similarly to the multi-class classification scenario, in each case, the model is trained five times on the randomly chosen subset of data to avoid the bias in the training data selection. The difference between the 2-class classification and the considered binary classification consists in the assumption about the fakes available at the training. The 2-class classification assumes that all types of fakes are available at the training stage whereas the binary classification assumes that only one type of fakes is available and the rest fakes are unknown. Obviously, the binary classification is more challenging and the results will highly depend on the type of fakes chosen for training. At the test stage all fakes are present for the classification. 

\begin{table*}[t!]
\begin{minipage}{\textwidth}
	\centering
	\renewcommand*{\arraystretch}{1.25}
	\caption{The classification error of the supervised binary classifier (in \%)\protect\footnote{Presented binary classification is close to the multi-class classification scenario with 2 classes considered in Section \ref{subsec_chap4:five class classification}. The difference in the obtained results is related to the presence of all types of fakes during the training in case of multi-class setup and randomly chosen training data.}.}
	\label{tab:supervised two class classification}.
	{\small
	\begin{tabular}{l|c|c|c|c|c} \hline
	%
	%
	Setup  & Originals ($P_{miss}$) & Fakes \#1 white ($P_{fa}$) & Fakes \#1 gray ($P_{fa}$) & Fakes \#2 white ($P_{fa}$) & Fakes \# 2 gray ($P_{fa}$) \\	\hline
	%
	Fakes \#1 white & 0 & 0 & 0.14 ($\pm$0.32) & 0 & 0 \\ \hline
	Fakes \#1 gray  & 0 & 0 & 0 & 0 & 0 \\ \hline
	Fakes \#2 white & 0 & 99.43 ($\pm$0.32) & 100 & 0 & 0 \\ \hline
	Fakes \# 2 gray & 0 & 99.29 ($\pm$0.5) & 99.86 ($\pm$0.32) & 0 & 0 \\ \hline
	\end{tabular}
	}
\end{minipage}	
\end{table*}
\begin{figure*}[t!]
	\centering
	%
	
	%
	\begin{subfigure}{0.24\textwidth}     
		\centering	
    \includegraphics[width=1.13\linewidth]{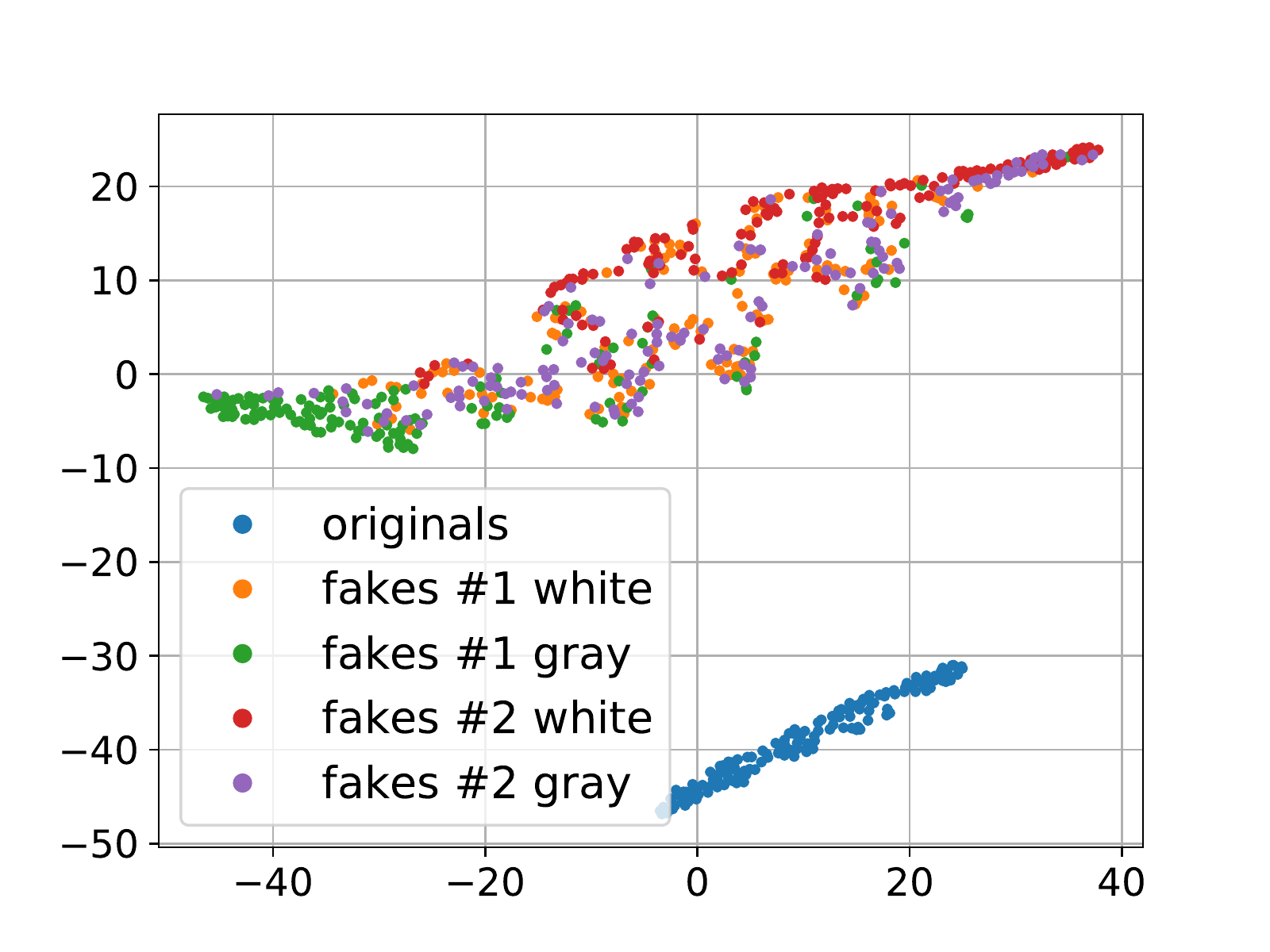}
        \caption{}
        \label{fig:tsne f1w}
	\end{subfigure}
	\begin{subfigure}{0.24\textwidth}
		\centering
    \includegraphics[width=1.13\linewidth]{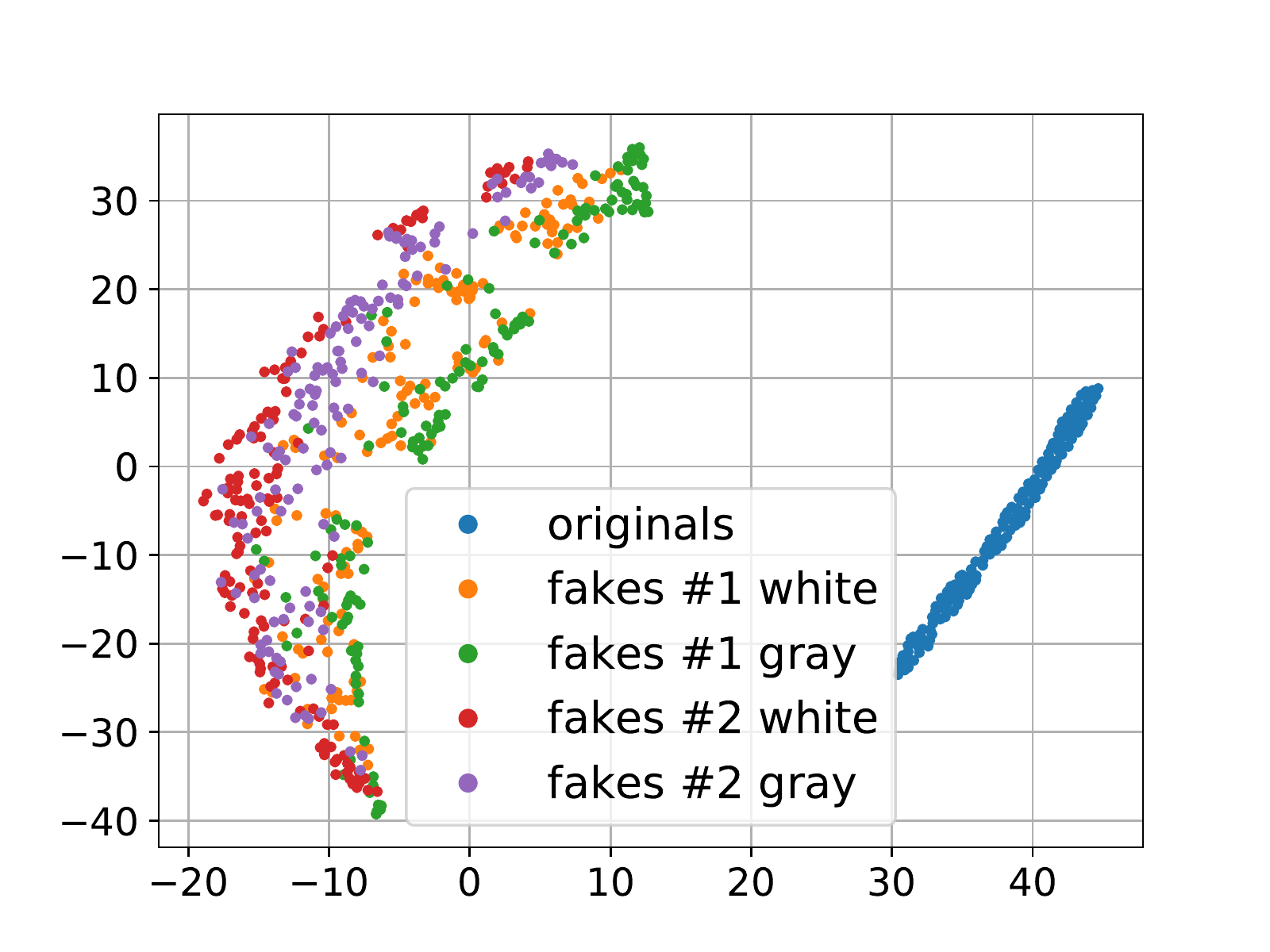}
        \caption{}
        \label{fig:tsne f1g}
	\end{subfigure}
	\begin{subfigure}{0.24\textwidth}     
		\centering	
    \includegraphics[width=1.13\linewidth]{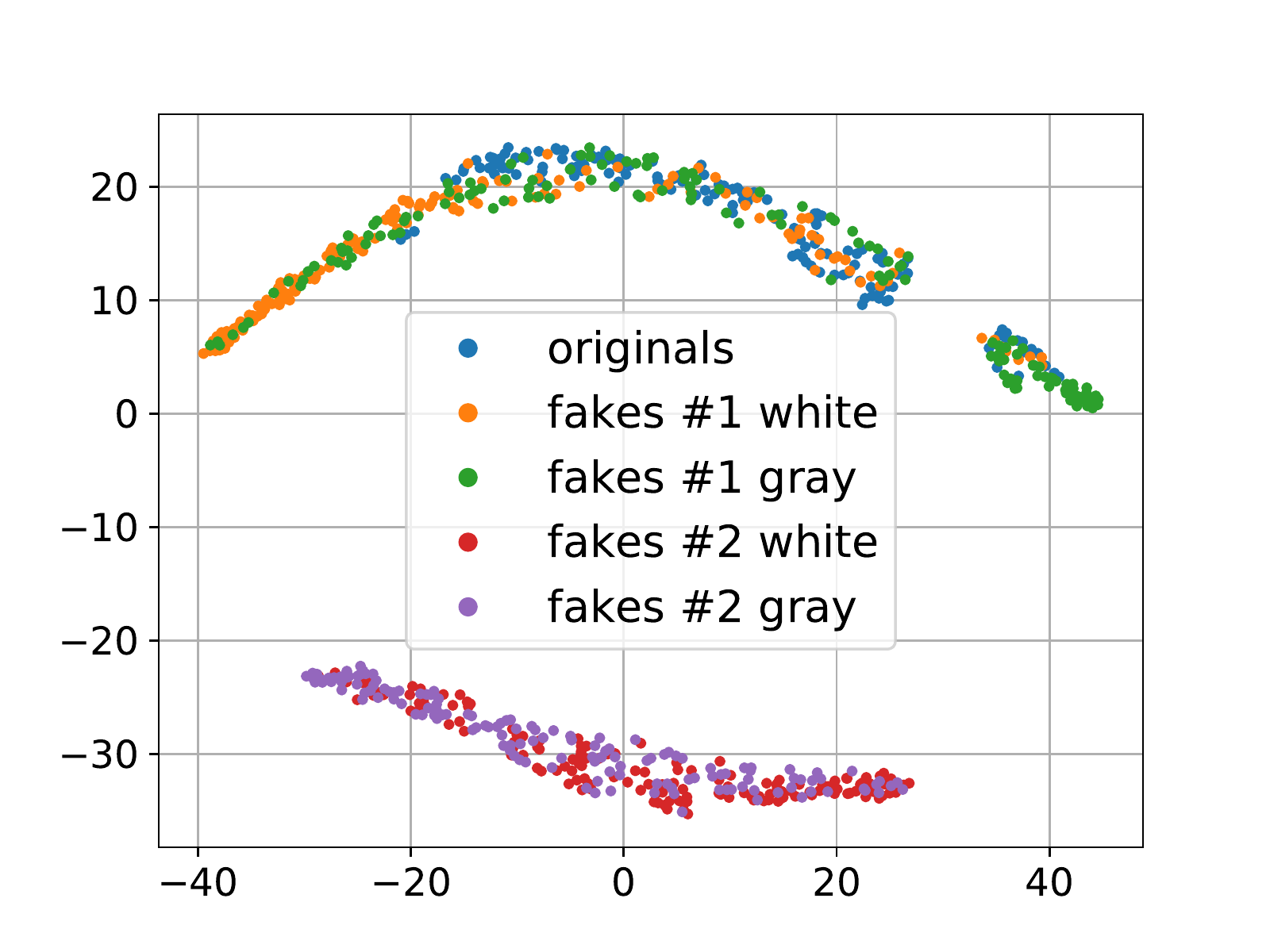}
        \caption{}
        \label{fig:tsne f2w}
	\end{subfigure}
	\begin{subfigure}{0.24\textwidth}
		\centering
    \includegraphics[width=1.13\linewidth]{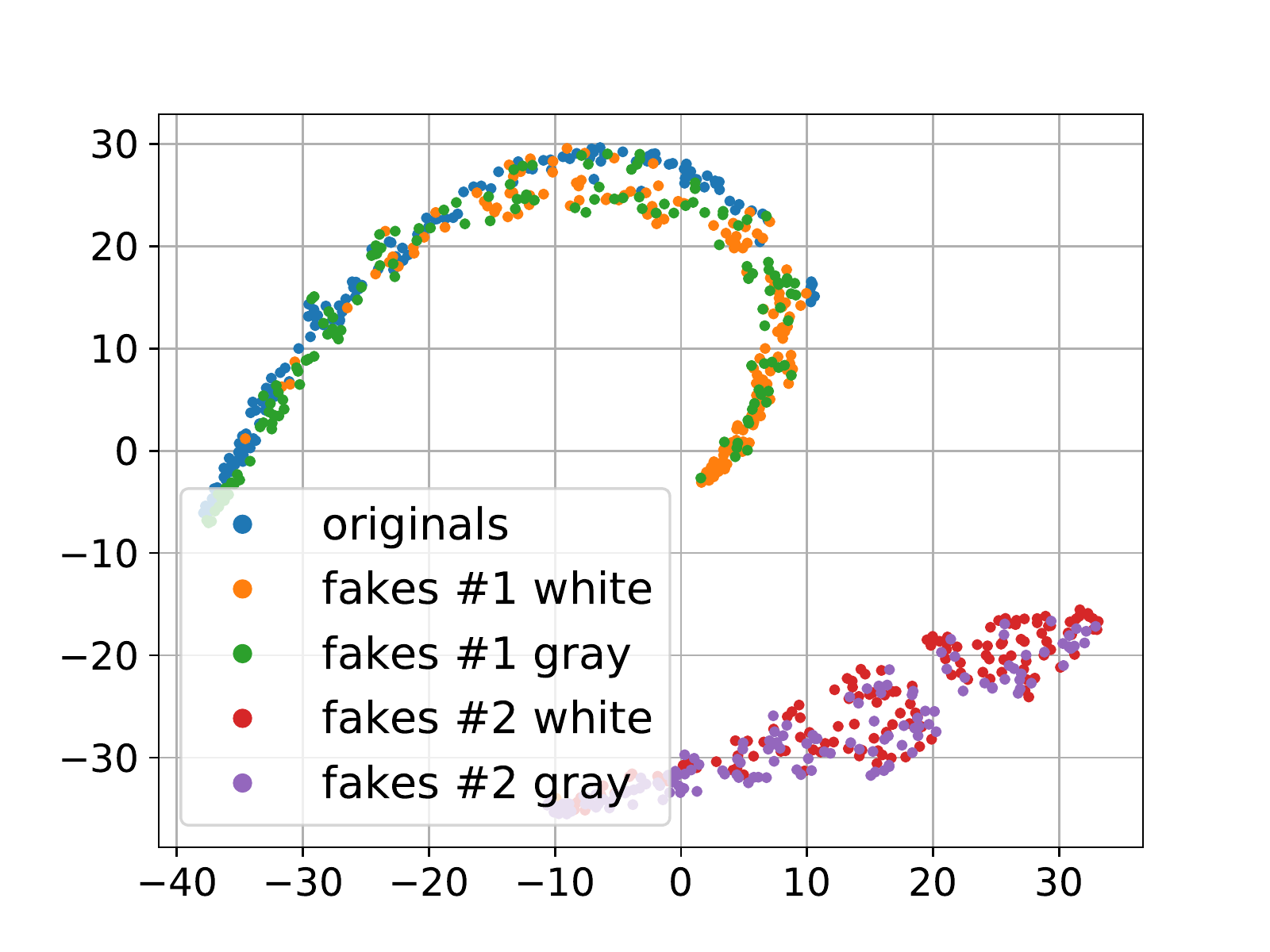}
        \caption{}
        \label{fig:tsne f2g}
	\end{subfigure}

    \caption{The latent space (the last layer before an activation function) t-SNE visualization of the supervised binary classifier trained on the originals and 
    (a) fakes \#1 white, (b) fakes \#1 gray, (c) fakes \#2 white, (d) fakes \#2 gray.}
    \label{fig:tsne binary classification}        
\end{figure*}		


The binary classification accuracy is evaluated  with respect to the probability of miss $P_{miss}$ and the probability of false acceptance $P_{fa}$ defined as: 
\begin{equation}
\left \{
\begin{array}{lll}
P_{fa}   & = & \textrm{Pr}\{ g_{\bphi}(\Y) = \c_1    \;|\; \mathcal{H}_0 \}, \\
P_{miss} & = & \textrm{Pr}\{ g_{\bphi}(\Y) \neq \c_1 \;|\; \mathcal{H}_1 \}, 

\end{array}
\right.
\label{eq:4}
\end{equation}
where $\c_1 = [1, 0]^\textrm{T}$ denotes a class of original codes, $\mathcal{H}_1$ corresponds to the hypothesis that the query $\y$ is an original code and $\mathcal{H}_0$ is the hypothesis that the query $\y$ is a fake code.

From the obtained results presented in Table \ref{tab:supervised two class  classification} one can note that both models trained on the originals and fakes \#1 provide high classification accuracy on all type of data, including the fakes \#2, unseen during the training. That is expected and can be explained by the fact that, as it is discussed in Section \ref{ch4_subsec:auth dataset}, the fakes \#1 are closer to the originals, while the fakes \#2 are the coarser copies of the original codes. In this regard, when the training is performed on the fakes \#2, no model is capable to distinguish the originals from the fakes \#1, unseen during the training. That is confirmed by the probability of false acceptance close to 100\%. Nevertheless, the models are capable to distinguish the originals from the fakes \#2 with 100\% accuracy. The t-SNE visualization of the latent space of each model illustrated in Fig. \ref{fig:tsne binary classification} confirms these observations. From Fig. \ref{fig:tsne f1w} and \ref{fig:tsne f1g} that present the latent space of models trained on the originals and the fakes \#1, one can see the good separability between the originals and fakes while all classes of fakes overlap. The latent space visualization of models trained on the originals and fakes \#2 illustrated in Fig. \ref{fig:tsne f2w} and \ref{fig:tsne f2g} shows the overlapping between the originals and the fakes \#1 preserving the fakes \#2 in well separable cluster.

\section{One-class classification}
\label{ch4_sec:one class classification}

\begin{table*}[t!]
\begin{minipage}{\textwidth}
	\centering
	\renewcommand*{\arraystretch}{1.25}
	\caption{The OC-SVM classification error in spatial domain (in \%)\protect\footnote{The python \textit{OneClassSVM} method from the sklearn package is used with the next training parameters: kernel="rbf"; gamma=0.1; nu=0.03 for the digital templates and nu=0.1 for the physical references.}.}
	\label{tab:oc-svm pearson vs hamming}
	{\small	
	\begin{tabular}{lccccc} \hline
		Train on & Originals ($P_{miss}$ ) & Fakes \#1 white ($P_{fa}$) & Fakes \#1 gray ($P_{fa}$) & Fakes \#2 white ($P_{fa}$) & Fakes \#2 gray ($P_{fa}$) \\ \hline
		\multicolumn{6}{l}{\textit{With respect to the digital templates:}} \\	
		\hspace{0.15cm} - grayscale $\x$  & 3.1 ($\pm$0.83) & 2.54 ($\pm$1.93) & 3.82 ($\pm$1.22) & 0 & 0 \\
		\hspace{0.15cm} - RGB $\x$        & 2.82 ($\pm$1.14) & 2.1 ($\pm$0.86) & 1.4 ($\pm$1.4) & 0 & 0  \\ \hline
		\multicolumn{6}{l}{\textit{With respect to the physical references:}} \\	
		\hspace{0.15cm} - grayscale $\x$ & 11.44 ($\pm$4.14) & 35.86 ($\pm$7.38) & 40.58 ($\pm$4.86) & 1.72 ($\pm$2.07) & 1.12 ($\pm$0.8) \\
		\hspace{0.15cm} - RGB $\x$       & 11.16 ($\pm$3.64) & 31.84 ($\pm$6.3) & 39.54 ($\pm$6.11) & 1.44 ($\pm$1.69) & 0.98 ($\pm$0.63) \\ \hline
	\end{tabular}
	}
\end{minipage}
\end{table*}    

\begin{figure*}[t!]
	\centering
	
	\begin{subfigure}{0.32\textwidth}
		\centering
        \includegraphics[width=0.9\linewidth,valign=t]{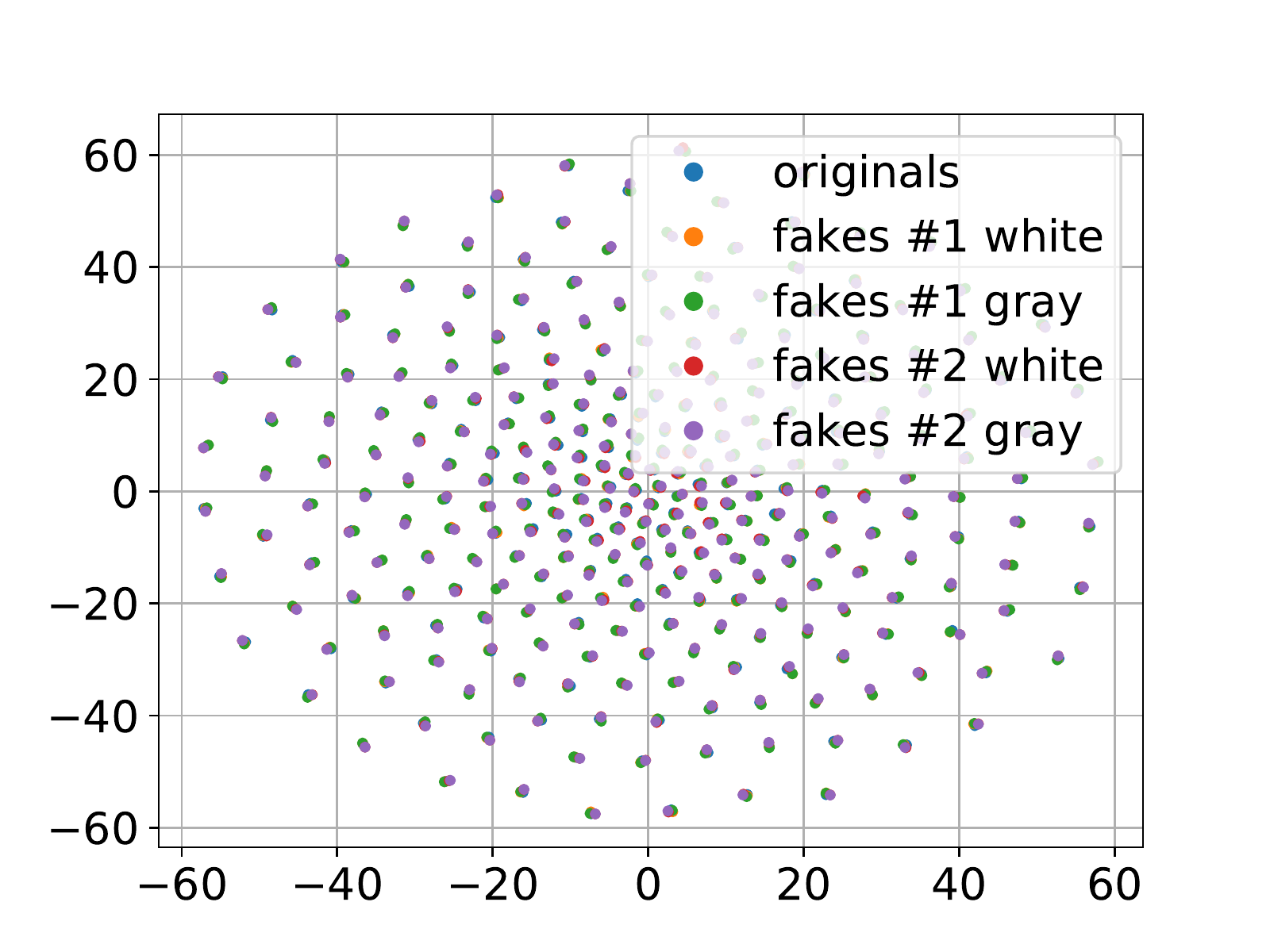}
	    \caption{}
	    \label{fig:tsne spatial domain direct rgb}
	\end{subfigure}		
	%
	%
	\begin{subfigure}{0.32\textwidth}
		\centering
        \includegraphics[width=0.9\linewidth,valign=t]{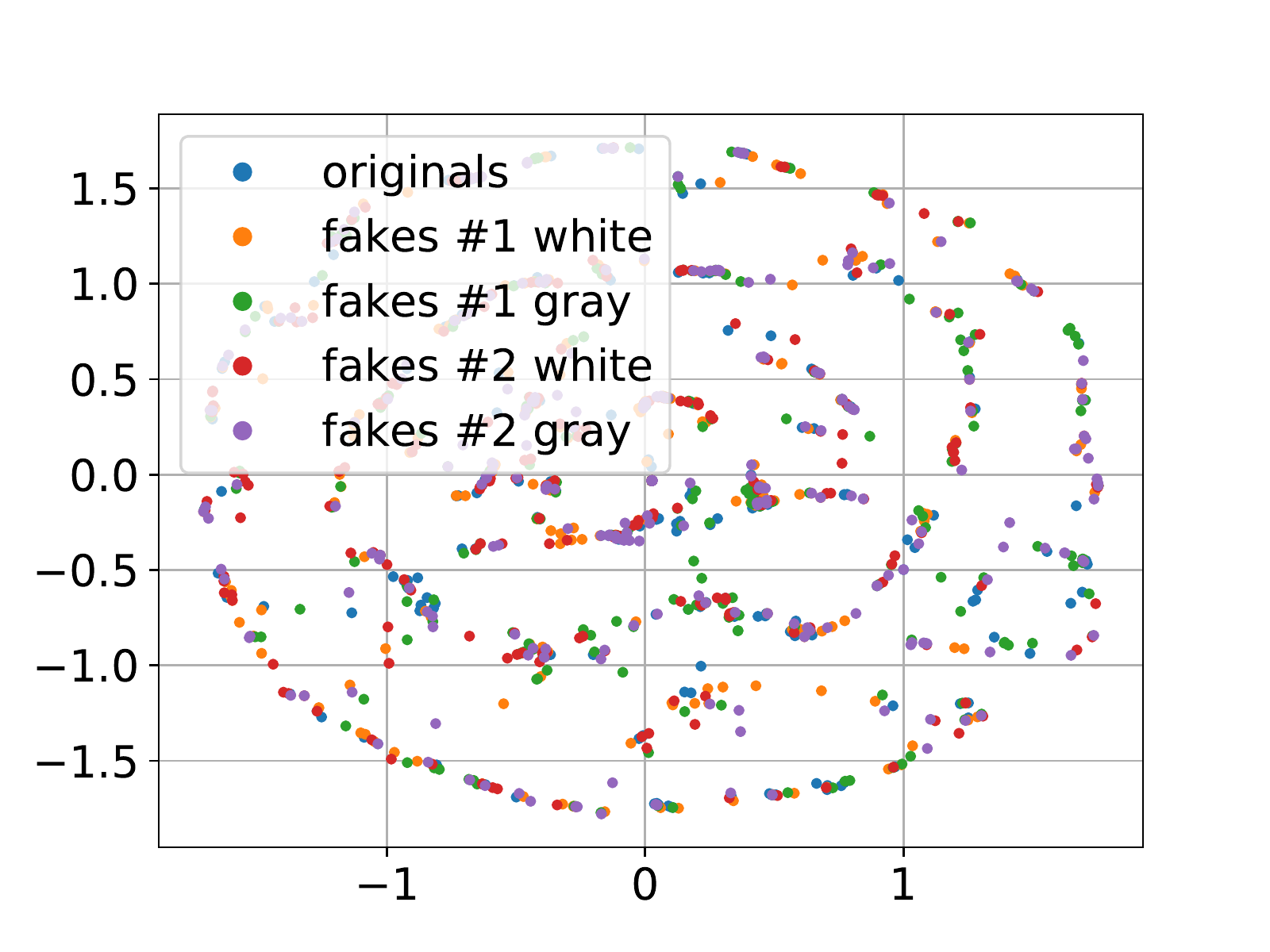}        
		\caption{}
		\label{fig:tsne spatial domain template diff}
	\end{subfigure}	
	%
	%
	\begin{subfigure}{0.32\textwidth}
		\centering
        \includegraphics[width=0.9\linewidth,valign=t]{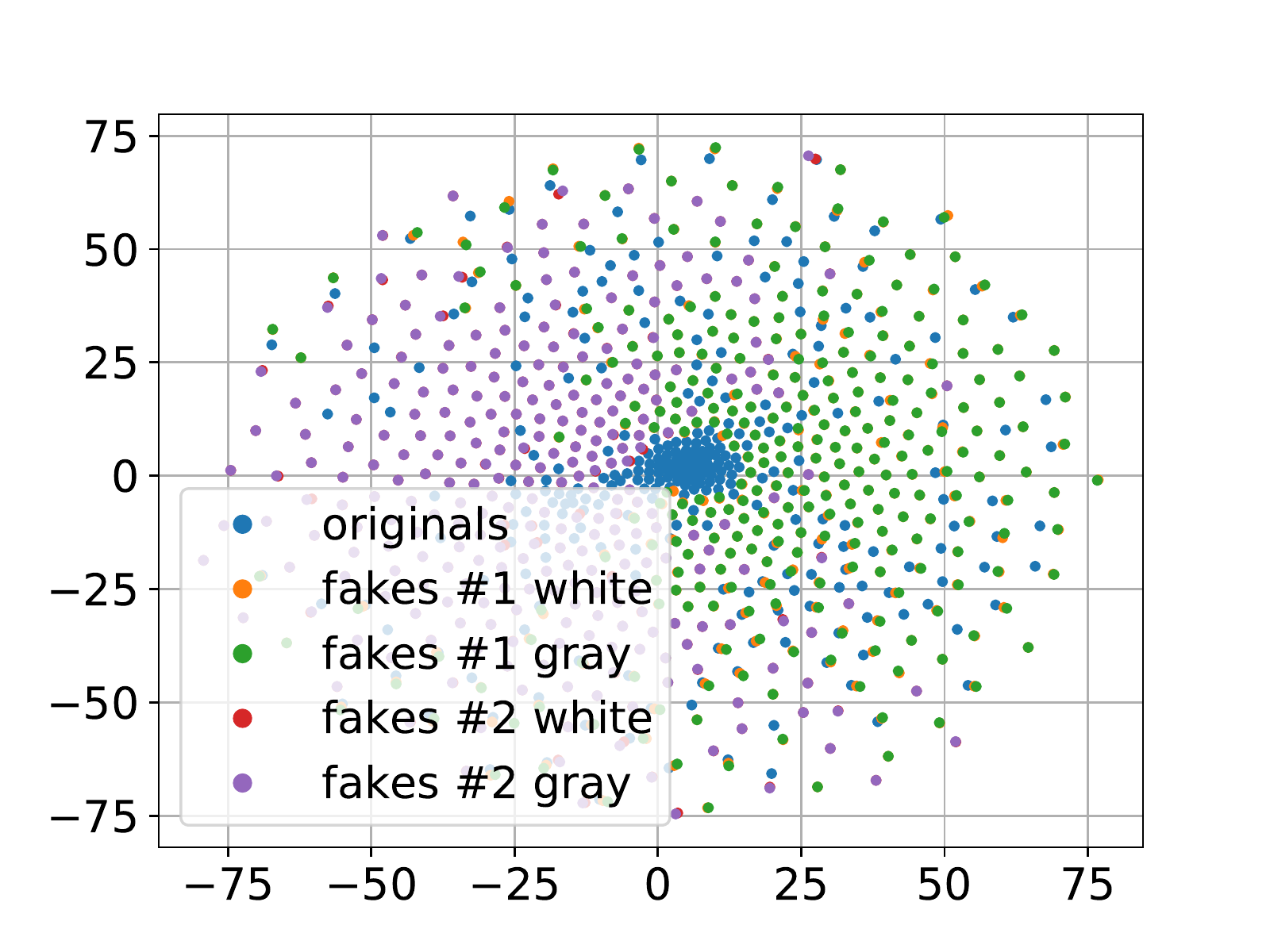}
		\caption{}
		\label{fig:tsne spatial domain printed reference diff}
	\end{subfigure}	
    \caption{The 2D t-SNE visualisation of the original and fake codes in the spatial domain (a horizontal axis denotes t-SNE dimension 1 and the t-SNE dimension 2 is on the vertical axis): (a) presents the direct RGB images' visualisation; (b) is based on the xor difference between the corresponding digital templates and printed codes binarized via a simple thresholding method with an optimal threshold determited individually for each printed code via the Otsu's method \cite{otsu1979threshold}; (c) visualizes the differences between the physical references and the corresponding printed original and fake codes.}
    \label{fig:tsne spatial domain}        
\end{figure*}


\begin{figure}[t!]
	\centering
	\begin{subfigure}{0.45\columnwidth}
		\centering
        \includegraphics[width=0.7\linewidth,valign=m]{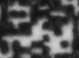}
        \caption{} 
        \label{fig:black_in_white_printed}
	\end{subfigure}		
	\hspace{0.2cm}
	\begin{subfigure}{0.45\columnwidth}
		\centering
        \includegraphics[width=0.7\linewidth,valign=m]{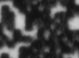}
        \caption{} 
        \label{fig:white_in_black_printed}
	\end{subfigure}		
    \caption{Examples of the dot gain effect: (a) a black symbol surrounded by white symbols increases its size but remains well detectable; (b) a white symbol surrounded by black symbols might disappear under strong dot gain.}
    \label{fig:dot gain examples}           
\end{figure}

\subsection{Spatial domain data analysis}
\label{ch4_subsec:spatial domain data analysis}

In Section \ref{sec_chap4: supervised classification} it is shown that according to results obtained for the Indigo mobile dataset the original and fake codes  are well separable in the latent space of the multi-class supervised classifier (Fig. \ref{fig:tsne five classes classification}). To answer the question how these data behave in the direct image domain (hereinafter also referred to as a \textit{spatial} domain), the 2D t-SNE visualizations of the data in the spatial domain are shown in Fig. \ref{fig:tsne spatial domain}. 

Fig. \ref{fig:tsne spatial domain direct rgb} shows the direct visualisation of the RGB images. One can note that the data do not form any clusters corresponding to originals or fakes. Instead, the data are allocated into small groups that are formed by the originals and fakes corresponding to the same digital template. Such a behavior is expectable and is explainable by the data nature. 

Fig. \ref{fig:tsne spatial domain template diff} demonstrates a visualization based on the xor difference between the digital templates and the corresponding printed codes binarized via a simple thresholding method with an optimal threshold determited individually for each printed code via the Otsu's method \cite{otsu1979threshold}. In general, one can observe a kind of rings that consist of the original and fakes but no clusters specific to the data types are observed. These rings are explainable by the fact that both originals and fakes can have bigger or smaller difference with the digital template due to the dot gain in the different group of black and white symbols as shown in Fig. \ref{fig:dot gain examples}: a white symbol surrounded by the black symbols results in a bigger binarization error, while the black symbol surrounded by the white symbols is more likely to survive after binarization. 

To better understand the role of the digital templates as a references, the Indigo mobile dataset was specially extended by the printed references (hereinafter also referred to as \textit{physical references}\footnote{The physical references correspond to the original codes acquired for the second time on the same equipment as the first case scenario. It assumes the probable presence of small geometrical (rotation) and illumination deviations between the original codes and corresponding physical references.}). It is easy to note the central dense cluster formed by the original codes (in blue) and two surrounding clusters from the fakes \#1 (mostly on the right-hand side) and fakes \#2 (mostly on the left-hand side) from Fig. \ref{fig:tsne spatial domain printed reference diff} that illustrates the t-SNE of the differences between the physical reference and the corresponding printed original and fake codes. Despite this, the overall mixing of individual samples from the different classes is quite significant. This indicates that the reliable direct spatial authentication might be complicated.

\begin{figure}[t!]
	\centering
	\begin{subfigure}{0.475\columnwidth}
		\centering
        \includegraphics[width=1\linewidth,valign=t]{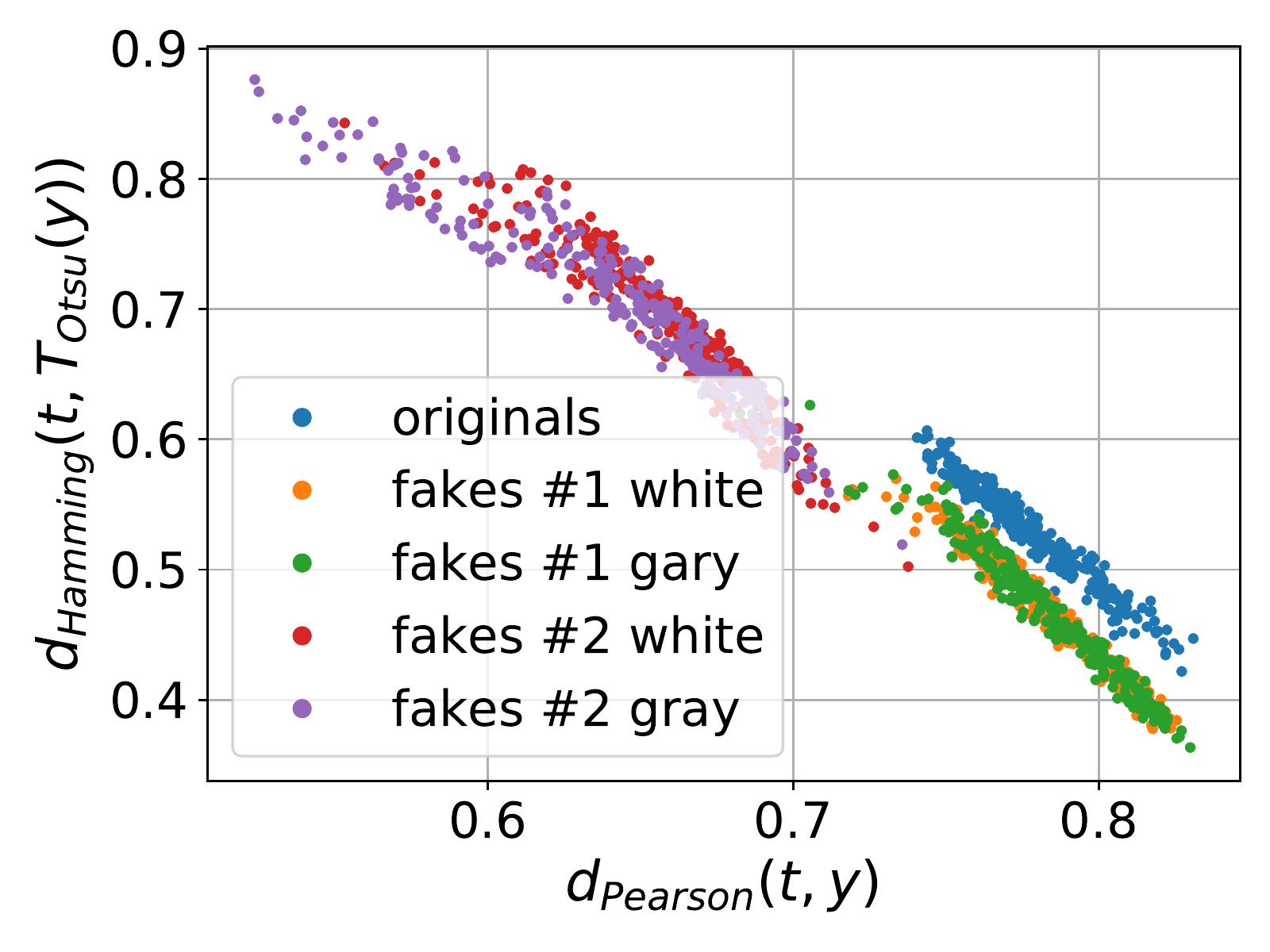}
	    \caption{With respect to the digital templates.}
	    \label{fig:spt domain digital pearson vs xor otsu main}
	\end{subfigure}		
	\hspace{0.2cm}
	\begin{subfigure}{0.475\columnwidth}
		\centering
        \includegraphics[width=1\linewidth,valign=t]{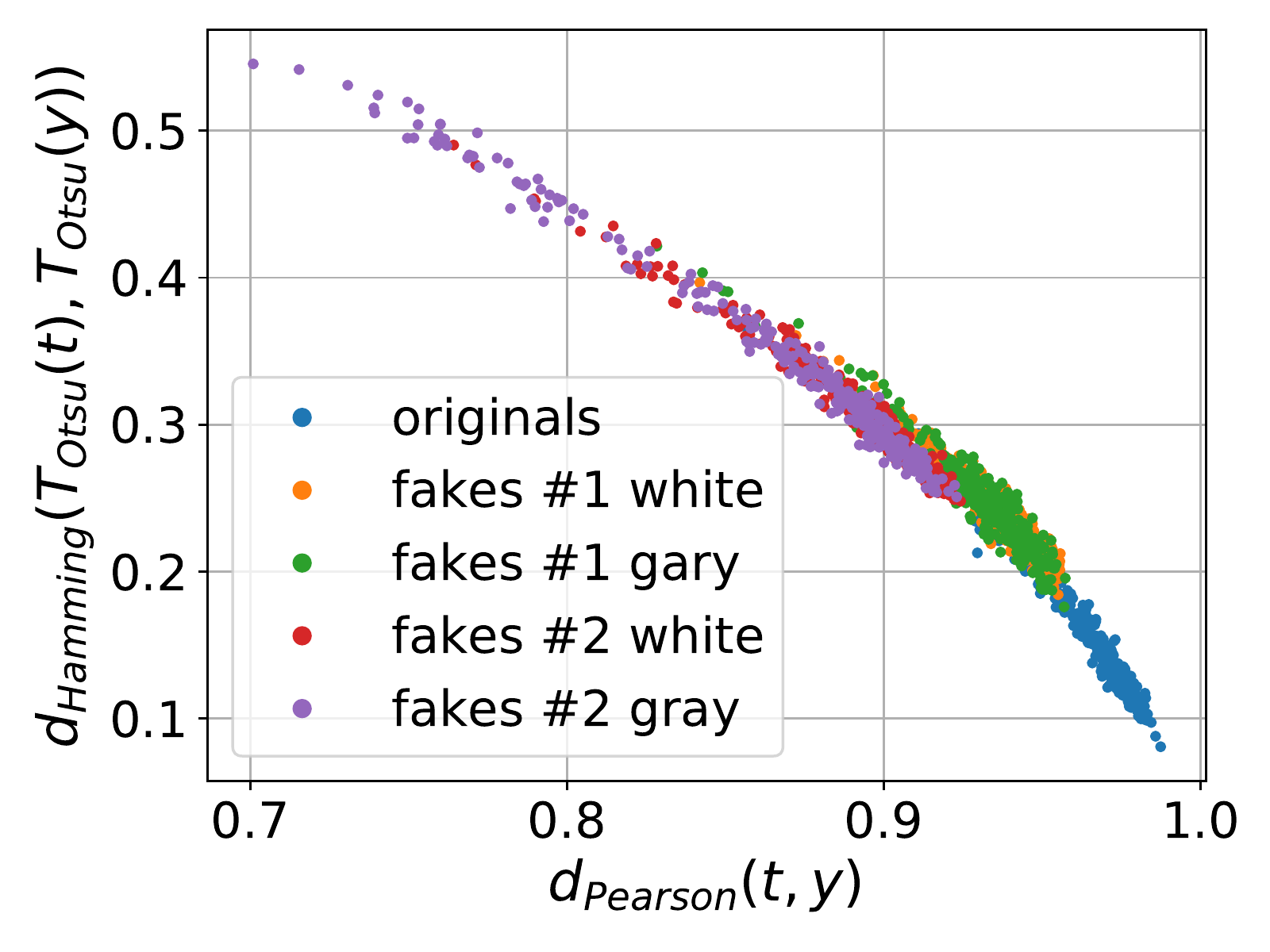}
	    \caption{With respect to the physical references.}
	    \label{fig:spt domain physical pearson vs xor otsu main}
	\end{subfigure}		
    \caption{The CDP separability in the 2D space of Pearson correlation (the horizontal axis) and Hamming distance (the vertical axis).}
    \label{fig:spatial domain digital vs physical two metric main}        
\end{figure}

As a next stage we performed the analysis of distances between the references (digital or physical) and the corresponding printed codes (original and fakes) in different metrics: $\ell_1$, $\ell_2$, Pearson correlation and Hamming distance. Whenever needed the binarization is applied via a simple thresholding with an optimal threshold determined individually for each code via the Otsu's method. The performed analysis demonstrates that besides some rare exceptions, it is impossible to separate the original and fake codes neither with respect to the digital template nor with respect to the physical reference based only on one metric. At the same time, the separability with respect to the two metrics is much better. The best two-metric separability we obtained is based on the Pearson correlation and Hamming distance between the printed codes and the corresponding digital or physical references  as shown in Fig. \ref{fig:spt domain digital pearson vs xor otsu main} - \ref{fig:spt domain physical pearson vs xor otsu main}. Encouraged by these results we apply the one-class support vector machines (OC-SVM) \cite{chen2001one} as a base-line approach for the classification of originals and fakes. 

To better understand the role of used reference and the influence of color information during the acquisition of black and white codes as opposed to their conversion to only grayscale images, the OC-SVM is applied with respect to four types of training data:
\begin{itemize}[noitemsep] 
	\item With respect to the digital templates on:
	\begin{itemize}[noitemsep,topsep=0pt] 
		\item the grayscale original codes $\x$;
		\item the RGB original codes $\x$.
	\end{itemize}
	\item With respect to the physical references on:
	\begin{itemize}[noitemsep,topsep=0pt] 
		\item the grayscale original codes $\x$;
		\item the RGB original codes $\x$.
	\end{itemize}	
\end{itemize}

\begin{figure*}[t!]
	\centering
	\begin{subfigure}{0.24\textwidth}
		\centering
        \includegraphics[width=0.95\linewidth,valign=t]{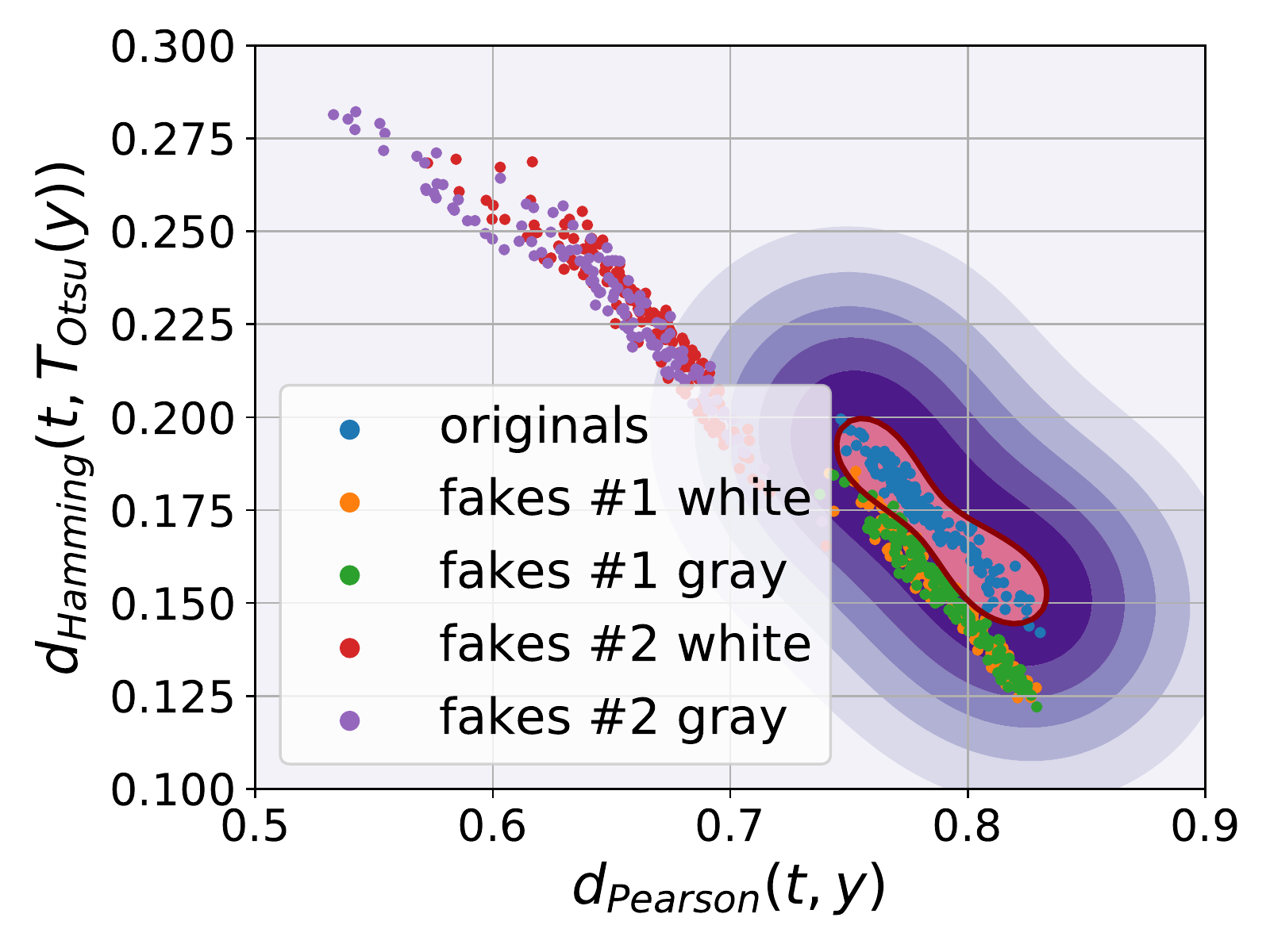}        
        \captionsetup{justification=centering}
	    \caption{With respect to the \textit{digital} templates: grayscale data.}
	    \label{fig:spatial domain digital gray one class svm}
	\end{subfigure}		
	\begin{subfigure}{0.24\textwidth}
		\centering
		\includegraphics[width=0.95\linewidth,valign=t]{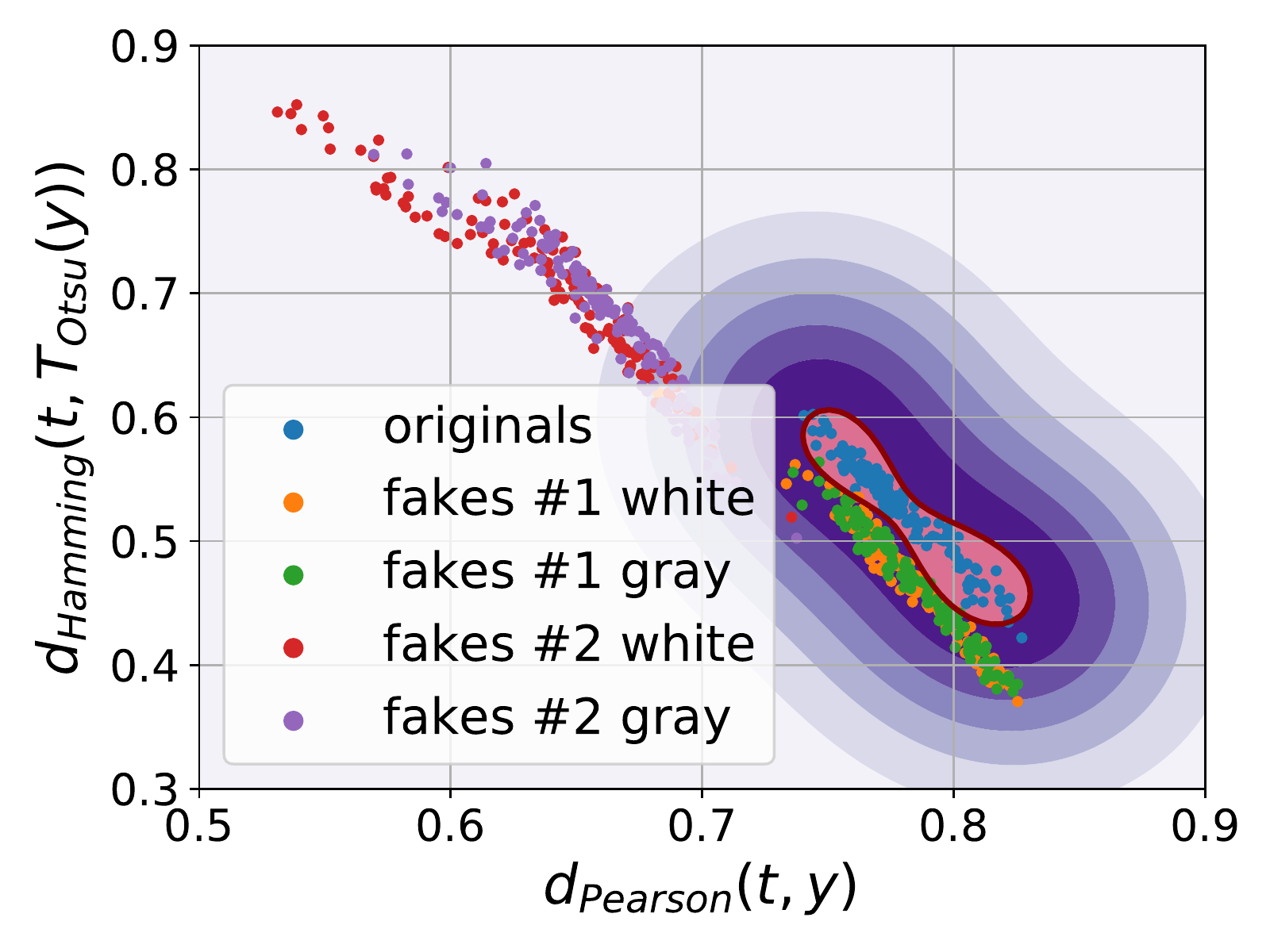}		
		\captionsetup{justification=centering}        
	    \caption{With respect to the \textit{digital} templates: RGB data.}
	    \label{fig:spatial domain digital rgb one class svm}
	\end{subfigure}		
	\begin{subfigure}{0.24\textwidth}
		\centering
        \includegraphics[width=0.95\linewidth,valign=t]{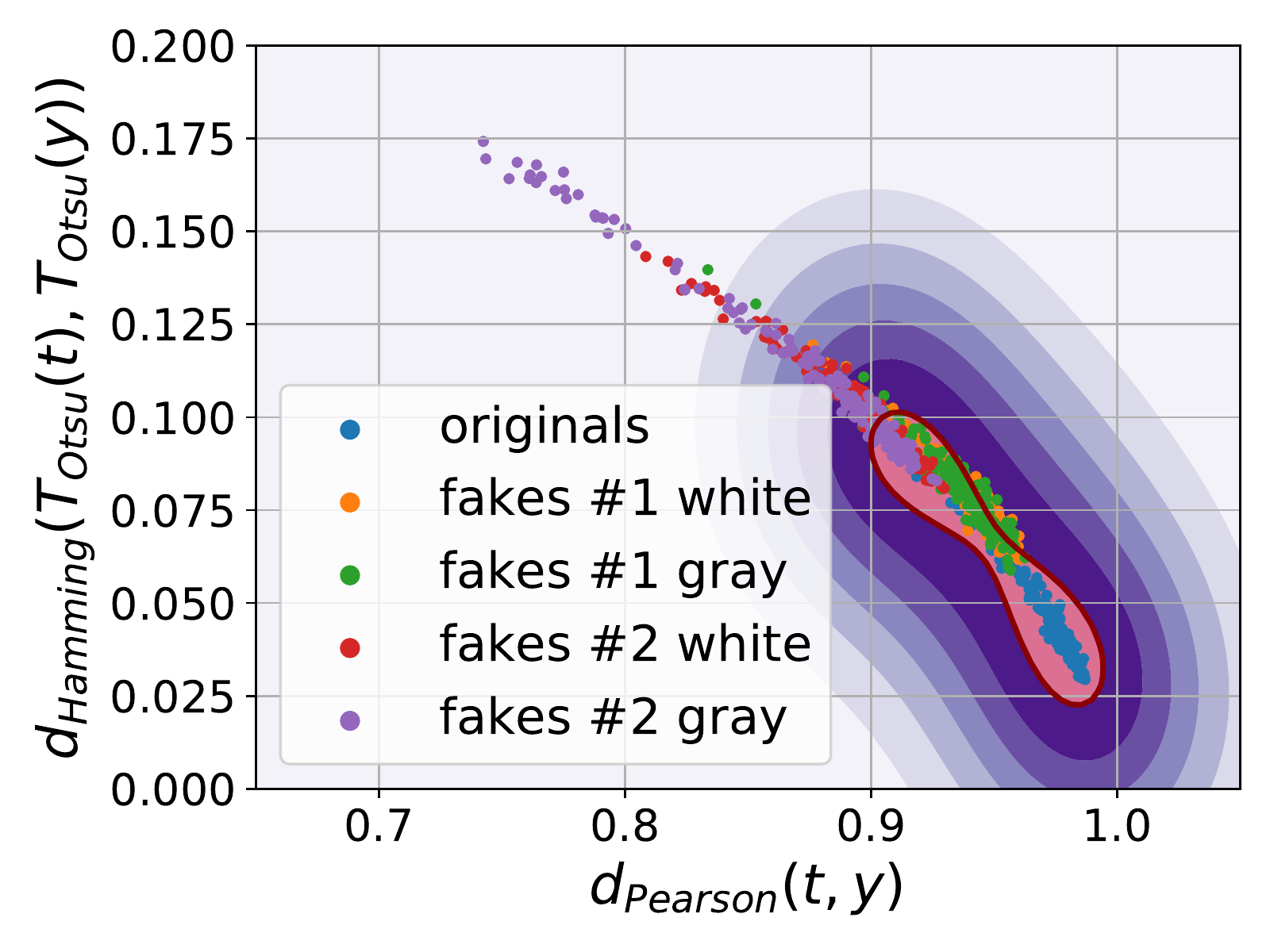}        
        \captionsetup{justification=centering}
	    \caption{With respect to the \textit{physical} references: grayscale data.}
	    \label{fig:spatial domain digital gray one class svm}
	\end{subfigure}		
	\begin{subfigure}{0.24\textwidth}
		\centering
        \includegraphics[width=0.95\linewidth,valign=t]{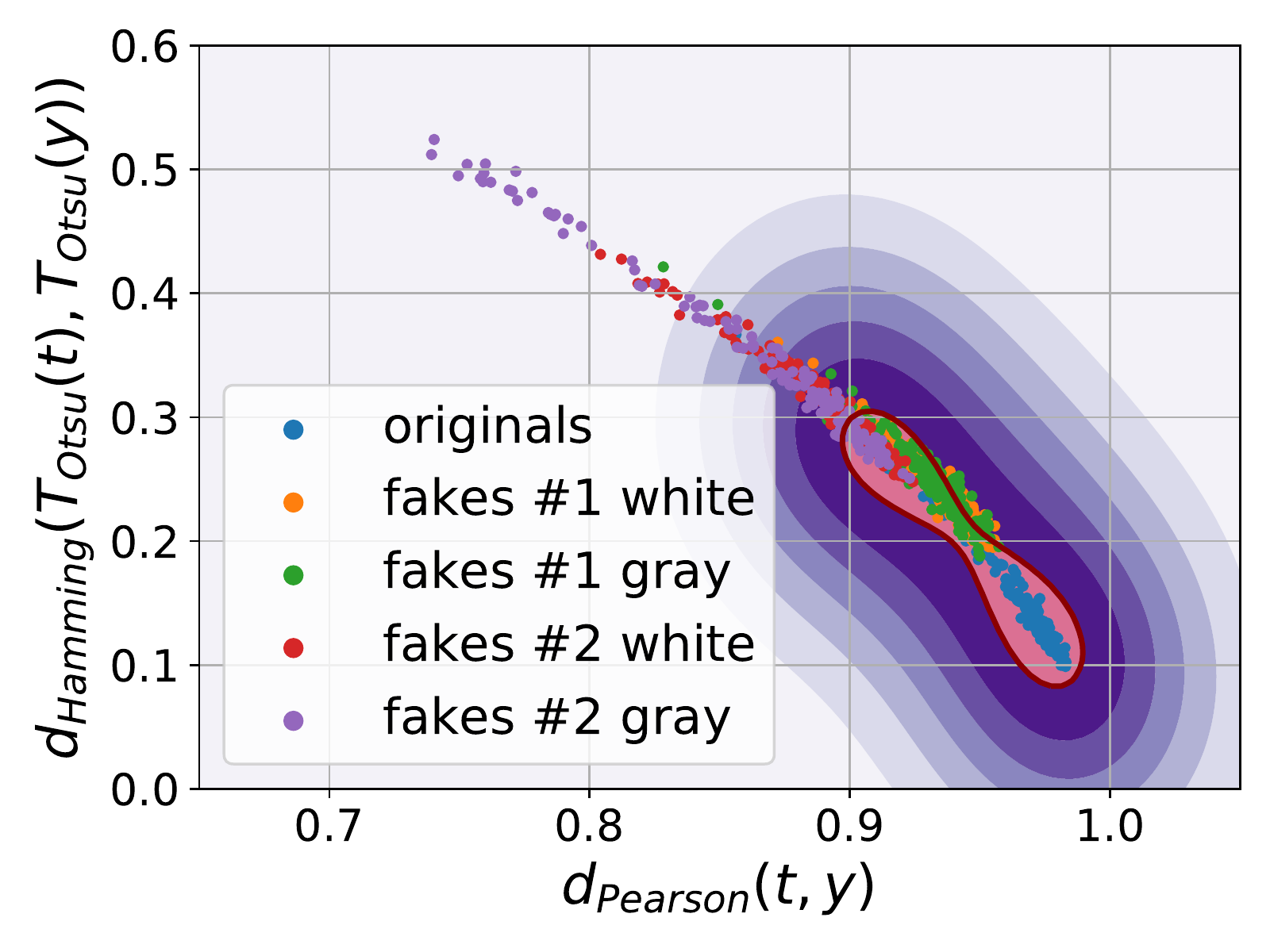}        
        \captionsetup{justification=centering}
	    \caption{With respect to the \textit{physical} references: RGB data.}
	    \label{fig:spatial domain digital rgb one class svm}
	\end{subfigure}		
    \caption{The decision boundaries of OC-SVM trained with respect to the Pearson correlation and Hamming distance between the reference (digital or physical) and the corresponding original printed codes.}
    \label{fig:spatial domain oc-svm decision boundaries}        
\end{figure*}


To avoid the bias in the training data selection, the OC-SVM was trained five times on randomly chosen original printed samples $\x$ and either digital templates or physical references. The OC-SVM was trained to minimize the $P_{miss}$ on the validation sub-set. The obtained classification error is represented in Table \ref{tab:oc-svm pearson vs hamming}. 
The visualisation of the OC-SVM decision boundaries is illustrated in Fig. \ref{fig:spatial domain oc-svm decision boundaries}. 

Analyzing the obtained results, at first, it should be pointed out that the OC-SVM classification error based on the $P_{miss}$ and $P_{fa}$ is relatively high. At the same time, two important conclusions can be done:

\begin{itemize}[noitemsep] 
	\item With respect to the chosen metrics the use of the digital templates is preferable than the printed references.
	\item Despite the visually grayscale nature of the CDP the authentication based on codes taken by the mobile phone in color mode is more efficient compared to the grayscale mode due to the fact that the different color channels have different sensitivity and due to the information loss while converting a three-channels color image into a single-channel grayscale one.
\end{itemize}


\subsection{Deep processing domain data analysis}
\label{ch4_subsec:one class classification from ib view}

\begin{figure}[t!]
    \centering
	\includegraphics[width=0.75\linewidth]{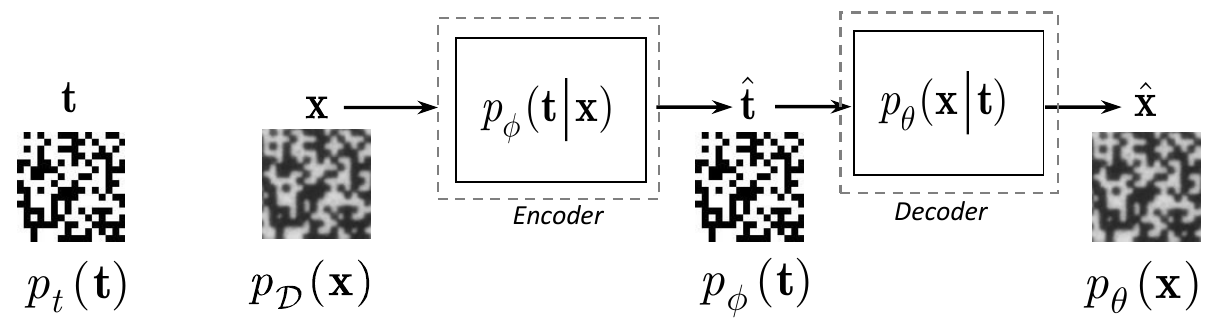}
	\caption{General scheme of a deep model that aims at estimating the digital templates $\hat{\t}$ from the original printed codes $\x$ with the following mapping of the estimated digital templates $\hat{\t}$ back to the printed codes $\hat{\x}$.}
	\label{fig:ib one class classification}	
\end{figure}

To further investigate the authentication performance, we consider an one-class classification based on the features extracted via DNN processing. In a particular case of the CDP authentication, where the  reference templates $\t$ are given, we consider a feature extractor based on a DNN auto-encoder model $\x \to \hat{\t} \to \hat{\x}$, where $\hat{\t}$ is considered as a latent space representation as shown in Fig. \ref{fig:ib one class classification}. The difference with a generic auto-encoder consists in the fact that the latent space is represented by a space of digital templates in contrast to some generic low-dimensional representation in a generic auto-encoder.

The loss-function for the considered feature extracting system is defined as:
%
\begin{equation} 
	\loss_{\textrm{One-class}}(\bphi, \btheta) =  - I_\bphi(\X; \T) - \beta I_{\bphi,\btheta}(\T; \X),
	\label{eq:deep feature extractor loss}
\end{equation}
where $\beta$ controls the relative importance of the two objectives.

The first mutual information term $I_\bphi(\X; \T)$ in (\ref{eq:deep feature extractor loss}) controls the mutual information between the estimate of template $\hat{\t}$ produced from $\x$ based on the mapper $p_\bphi(\t|\x)$ and original template $\t$ and is defined as:
\begin{equation}
\begin{aligned} 
	 I_\bphi(\X; \T)  
	 & = \mathbb{E}_{p(\x,\t)}  \left[  \log \frac{p(\x,\t)}{\pdx {p_t({\t})}}  \right] \\
	 & = \mathbb{E}_{p(\x,\t)}  \left[  \log \frac{\pdx p_\bphi(\t|\x)}{\pdx {p_t({\t})}}  \right] \\
	 & =\mathbb{E}_{p(\x,\t)}  \left[  \log \frac{p_\bphi(\t|\x)}{p_t(\t)}  \right].
\end{aligned} 	 
\label{ib-occ second term direct decomposition}
\end{equation}
%

%
According to \cite{Svolos2020entropy}, the variational decomposition is applied to decompose (\ref{ib-occ second term direct decomposition}) into a form suitable for the practical calculations:
\begin{equation}
\begin{aligned} 
	 I_\bphi(\X; \T)  
	 & = \mathbb{E}_{p(\x,\t)}  \left[  \log \frac{p_\bphi(\t|\x)}{p_t(\t)} \frac{p_\bphi(\t)}{p_\bphi(\t)} \right] \\
	 & = - \mathbb{E}_{p_t(t)} \left[   \log \frac{p_t(\t)}{p_\bphi(\t)}  \right] - \mathbb{E}_{p_t(\t)} \left[ \log p_\bphi(\t)\right] \\
	 & \;\;\;\; + \mathbb{E}_{p_\D(\x)} \left[ \mathbb{E}_{p_\bphi(\t|\x)} \left[ \log p_\bphi(\t|\x) \right] \right],
	 %
\end{aligned} 	 
\end{equation}
where $D_{\mathrm{KL}}\left( p_t(\t) \| p_\bphi(\t) \right) = \mathbb{E}_{p_t(t)} \left[   \log \frac{p_t(\t)}{p_\bphi(\t)}  \right]$ is a Kullback–Leibler divergences between the true $p_t(\t)$ and the posterior $p_\bphi(\t)$. $H(p_t(\t), p_\bphi(\t)) = - \mathbb{E}_{p_t(\t)} \left[ \log p_\bphi(\t)\right]$ is a cross-entropy. 

Taking into account that the cross-entropy $H(p_t(\t), p_\bphi(\t)) \ge 0$, we get $I_\bphi(\X; \T) \ge I_\bphi^L(\X; \T)$, where:
\begin{equation} 
\begin{aligned} 
    I_\bphi^L(\X; \T) 
    & \triangleq \underbrace{\mathbb{E}_{p_\D(\x)} \left[ \mathbb{E}_{p_\bphi(\t|\x)} \left[ \log p_\bphi(\t|\x) \right] \right]}_\text{$\Dtt$} \\
    & \;\;\; - \underbrace{D_{\mathrm{KL}}\left( p_t(\t) \| p_\bphi(\t) \right)}_\text{$\Dt$}.
\end{aligned}     
\label{eq:first term}  
\end{equation}

\begin{figure}[t!]
	\centering
    \includegraphics[width=1.03\columnwidth]{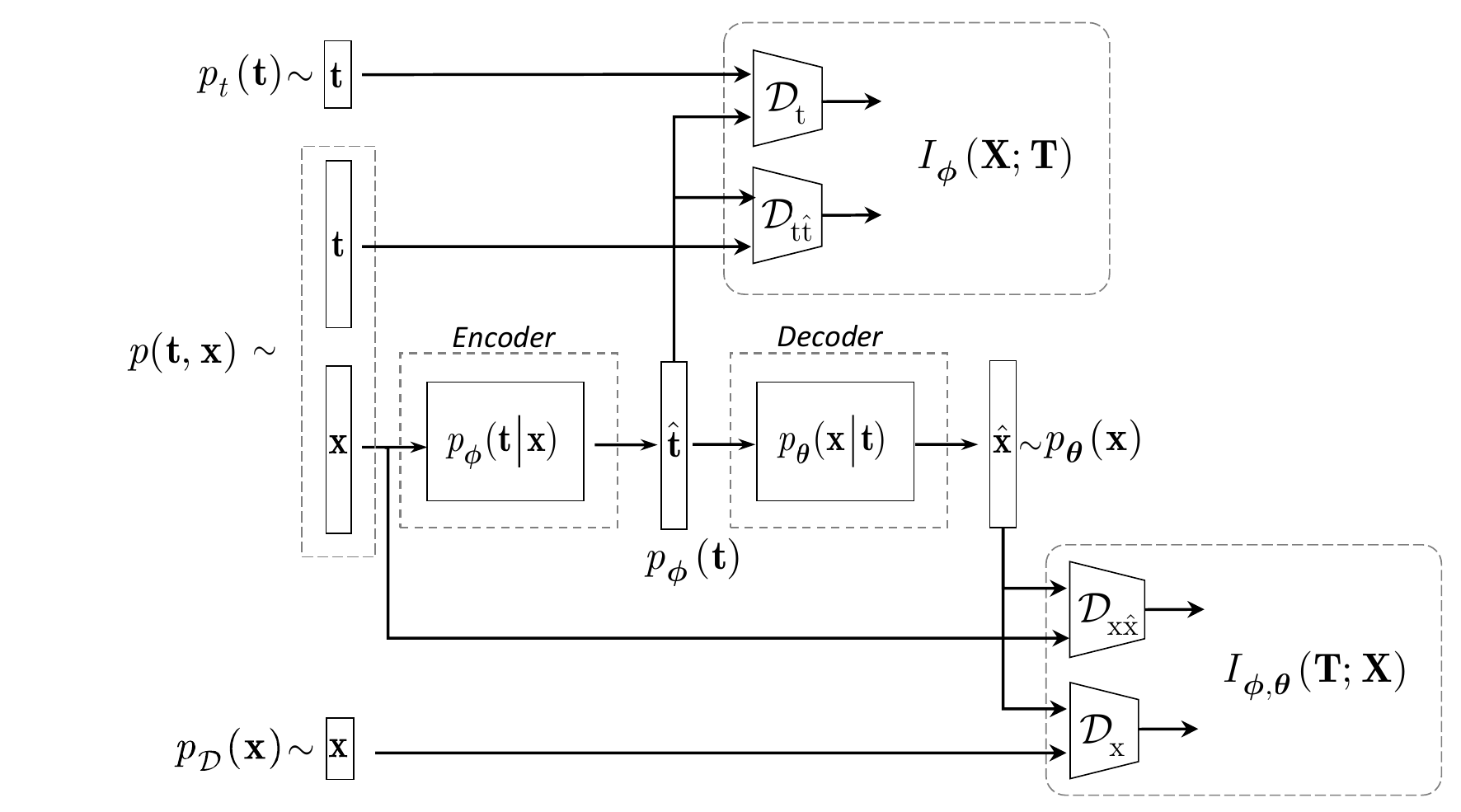}
    \caption{The feature extraction for the one-class classification based on the estimation of the reference templates  via $\Dtt$ and $\Dt$ and the printed codes via $\Dxx$ and $\Dx$ terms.}
    \label{fig:dtt dt dxx dx hc scheme}
\end{figure}


 The second mutual information term in (\ref{eq:deep feature extractor loss}) determined as $I_{\bphi,\btheta}(\T;\X) = \mathbb{E}_{p_\D(\x)} \left[ \mathbb{E}_{p_\bphi(\t|\x)} \left[ \log \frac{p_\btheta(\x|\t)}{p_\D(\x)} \right] \right]$  can  be  decomposed  and bounded  in  a  way  similar  to  the  first  term: $I_{\bphi,\btheta}(\T;\X) \geq I^L_{\bphi, \btheta}(\T;\X)$, where:
%

\begin{equation}
\begin{aligned} 
I^L_{\bphi,\btheta}(\T;\X) 
    & \triangleq \underbrace{\mathbb{E}_{p_\D(\x)}  \left[ \mathbb{E}_{p_{\bphi}(\t|\x)} \left[  \log p_{\btheta}(\x| \t) \right]\right]}_\text{$\Dxx$} \\
    & \;\;\; - \underbrace{D_{\mathrm{KL}}\left(p_\D(\x ) \| p_{\btheta}(\x)\right)}_\text{$\Dx$}.
\end{aligned} 
\label{eq:second term}  
\end{equation}

{\bf Remark:} The term $\Dt$ in (\ref{eq:first term}) and the term $\Dx$ in (\ref{eq:second term}) can be implemented based on the density ratio estimation \cite{GoodfellowGAN}. The terms $\Dtt$ and $\Dxx$ can be defined explicitly using Gaussian or Laplacian priors. In the Gaussian case, one can define $p_\bphi(\t|\x)  \propto \exp(-\lambda_1\| \t - g_{\bphi}(\x)\|_2)$ and $p_\btheta(\x|\t)  \propto \exp(-\lambda_2\| \x - f_{\btheta}(\t)\|_2)$ with the scale parameters $\lambda_1$ and $\lambda_2$, which lead to $\ell_2$-norm, and $g_{\bphi}(\x)$ denotes the encoder and $f_{\btheta}$ denotes the decoder. It also corresponds to the model $\t = g_{\bphi}(\x) + {\bf e}_\textrm{x}$ and $\x = f_{\btheta}(\t) + {\bf e}_\textrm{t}$, where ${\bf e}_\textrm{x}$ and ${\bf e}_\textrm{t}$ are the corresponding reconstruction error vectors following the Gaussian pdf. 

Thus, the equation (\ref{eq:first term}) reduces to:
\begin{equation}
\begin{aligned} 
    I^L_\bphi(\X;\T)  
    & = \underbrace{ - \lambda_1 \mathbb{E}_{p_\D(\x)}  \left[ \mathbb{E}_{p_{\bphi}(\t|\x)} \left[\| \t - g_{\bphi}(\x)\|_2 \right]\right]}_\text{$\Dtt$} \\
    & \;\;\; - \underbrace{D_{\mathrm{KL}}\left(p_t(\t ) \| p_{\bphi}(\t)\right)}_\text{$\Dt$},
\end{aligned}
\end{equation}

and (\ref{eq:second term}) reduces to:

\begin{equation}
\begin{aligned} 
    I^L_{\bphi,\btheta}(\T;\X) 
    & \triangleq \underbrace{- \lambda_2\mathbb{E}_{p_\D(\x)}  \left[ \mathbb{E}_{p_{\bphi}(\t|\x)} \left[  \| \x - f_{\btheta}(\t)\|_2 \right]\right]}_\text{$\Dxx$} \\
    & \;\;\; - \underbrace{D_{\mathrm{KL}}\left(p_\D(\x ) \| p_{\btheta}(\x)\right)}_\text{$\Dx$},
\end{aligned} 
\end{equation}


%


\begin{table*}[t!]
\begin{minipage}{\textwidth}
	\centering
	\renewcommand*{\arraystretch}{1.25}
	\caption{The OC-SVM classification error in deep processing domain (in \%)\protect\footnote{The python \textit{OneClassSVM} method from the sklearn package is used with the following training parameters: kernel="rbf"; gamma=0.1; nu=0.0005.}.}
	\label{tab:ib oc classification results}
	{\small
	\begin{tabular}{lcccccc} \hline
		%
		\multirow{2}{*}{Model} & & Originals & Fakes \#1 white & Fakes \#1 gray & Fakes \#2 white & Fakes \#2 gray \\
		& & $P_{miss}$ & $P_{fa}$ & $P_{fa}$ & $P_{fa}$ & $P_{fa}$ \\[0.1cm] \hline 
		\multicolumn{7}{c}{\textit{Based on the equation (\ref{ch4_eq:one metric decision})}} \\
		$\loss_\textrm{One-class}^1: - \Dtt$ & & 0 & 6.38 ($\pm$2.4) & 8.23 ($\pm$2.95) & 0 & 0 \\ 
		$\loss_\textrm{One-class}^2: - \Dtt +  \Dt$ & & 0 & 6.81 ($\pm$1.63) & 7.09 ($\pm$2.4) & 0 & 0 \\
		$\loss_\textrm{One-class}^3: - \Dtt - \beta \Dxx$ & & 0 & 1.56 ($\pm$0.32) & 0.99 ($\pm$0.81) & 0 & 0 \\
		$\loss_\textrm{One-class}^4: - \Dtt +  \Dt - \beta \Dxx + \beta \Dx$ & & 0 & 2.41 ($\pm$1.38) & 2.13 ($\pm$1.59) & 0 & 0 \\ \hline
		\multicolumn{7}{c}{\textit{Based on the equation (\ref{ch4_eq:two metric decision})}} \\
		$\loss_\textrm{One-class}^3: - \Dtt - \beta \Dxx$ & & 0 & 0.28 ($\pm$0.64) & 0 & 0 & 0 \\
		$\loss_\textrm{One-class}^4: - \Dtt +  \Dt - \beta \Dxx + \beta \Dx$ & & 0.57 ($\pm$1.27) & 0 & 0.14 ($\pm$0.32) & 0 & 0 \\ \hline
		\multicolumn{7}{c}{\textit{Based on the OC-SVM}} \\
		$\loss_\textrm{One-class}^3: - \Dtt - \beta \Dxx$ & & 0.28 ($\pm$0.39) & 0 & 0 & 0 & 0 \\
		$\loss_\textrm{One-class}^4: - \Dtt +  \Dt - \beta \Dxx + \beta \Dx$ & & 0.14 ($\pm$0.32) & 0 & 0 & 0 & 0 \\ \hline
	\end{tabular}
	}
\end{minipage}
\end{table*}    

The final optimization problem schematically shown in Fig. \ref{fig:dtt dt dxx dx hc scheme} is:
\begin{equation} 
\begin{aligned}
(\hat{\bphi}, \hat{\btheta}) & = \argmin_{\bphi, \btheta}  
\loss_{\textrm{One-class}}^{L}(\bphi, \btheta) \\
& = \argmin_{\bphi, \btheta}  - (\Dtt - \Dt) - \beta (\Dxx - \Dx).
\end{aligned}
\label{eq:IB_final_optimization}
\end{equation}
%
where:
\begin{equation}
\begin{aligned}
\Dtt & \triangleq \mathbb{E}_{p_\D(\x)} \left[ \mathbb{E}_{p_\bphi(\t|\x)} \left[ \log p_\bphi(\t|\x) \right] \right],
\\ 
\Dt & \triangleq D_{\mathrm{KL}}\left( p_t(\t) \| p_\bphi(\t) \right),
\\ 
\Dxx & \triangleq \mathbb{E}_{p_\D(\x)}  \left[ \mathbb{E}_{p_{\bphi}(\t|\x)} \left[  \log p_{\btheta}(\x| \t) \right]\right],
\\
\Dx & \triangleq  D_{\mathrm{KL}}\left(p_\D(\x) \| p_{\btheta}(\x)\right).
\end{aligned}
\label{eq:ib-occ all terms}  
\end{equation}

In practice we considered four basic scenarios of features extractors for the one-class classification:
\begin{enumerate}
	\item The reference templates estimation based on the term $\Dtt$:
	\begin{equation}
		\loss_\textrm{One-class}^1(\bphi, \btheta) = -\Dtt.
		\label{ch4_eq:oc classification dtt}
	\end{equation}
	\item The reference templates estimation based on the terms $\Dtt$ and $\Dt$:
	\begin{equation}
		\loss_\textrm{One-class}^2(\bphi, \btheta) = -\Dtt + \Dt.
		\label{ch4_eq:oc classification dtt dt}		
	\end{equation}	
	\item The estimation of the reference templates and the printed codes based on terms $\Dtt$ and $\Dxx$:
	\begin{equation}
		\loss_\textrm{One-class}^3(\bphi, \btheta) = -\Dtt - \beta  \Dxx. 
		\label{ch4_eq:oc classification dtt dxx}		
	\end{equation}
	\item The estimation of the reference templates and the printed codes based on terms $\Dtt$, $\Dt$, $\Dxx$ and $\Dx$:
	\begin{equation}
		\loss_\textrm{One-class}^4(\bphi, \btheta) = -\Dtt + \Dt - \beta \Dxx + \beta \Dx.
		\label{ch4_eq:oc classification dtt dt dxx dx}		
	\end{equation}
\end{enumerate}

\begin{figure}[t!]
	\centering
    \includegraphics[width=1\columnwidth]{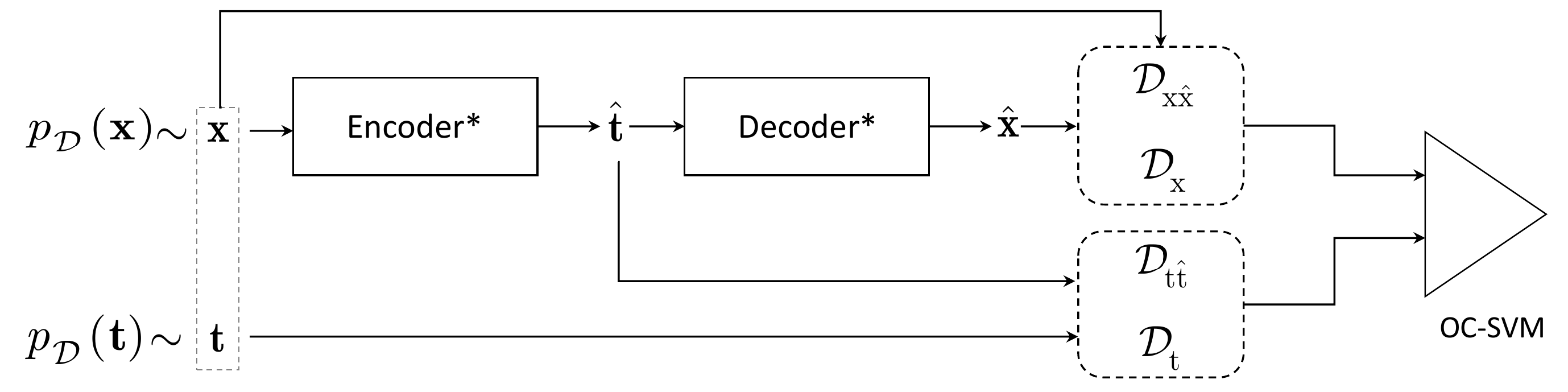}
    \caption{The one-class classification training procedure: the encoder and decoder parts of the auto-encoder model shown in Fig. \ref{fig:dtt dt dxx dx hc scheme} are pre-trained and fixed (as indicated by a "*"); the OC-SVM is trained on the outputs of $\Dtt$ and $\Dt$ terms that are the results of $I^L_{\bphi}(\X;\T)$ decomposition and the $\Dxx$ and $\Dx$ terms that are the results of $I^L_{\bphi,\btheta}(\T;\X)$ decomposition.}
    \label{fig:dtt dt dxx dx oc classification scheme}
\end{figure}

In general case, to be comparable with the one-class classification in the spatial domain discussed in Section \ref{ch4_subsec:spatial domain data analysis}, the one-class classification model based on the OC-SVM is used. 

The OC-SVM training procedure shown in Fig. \ref{fig:dtt dt dxx dx oc classification scheme} uses the pre-trained and fixed encoder and decoder parts of the auto-encoder model that serves as a features extractor. As an input the OC-SVM might take different combinations of outputs of four main terms: $\Dtt$, $\Dt$,  $\Dxx$ and $\Dx$. The exact scenarios are discussed in Sections \ref{ch4_subsubsec:dtt} - \ref{ch4_subsubsec:dtt dt dxx dx} below.


\subsection{Experimental results}

\begin{figure}[t!]
    \centering
    \includegraphics[width=0.75\linewidth]{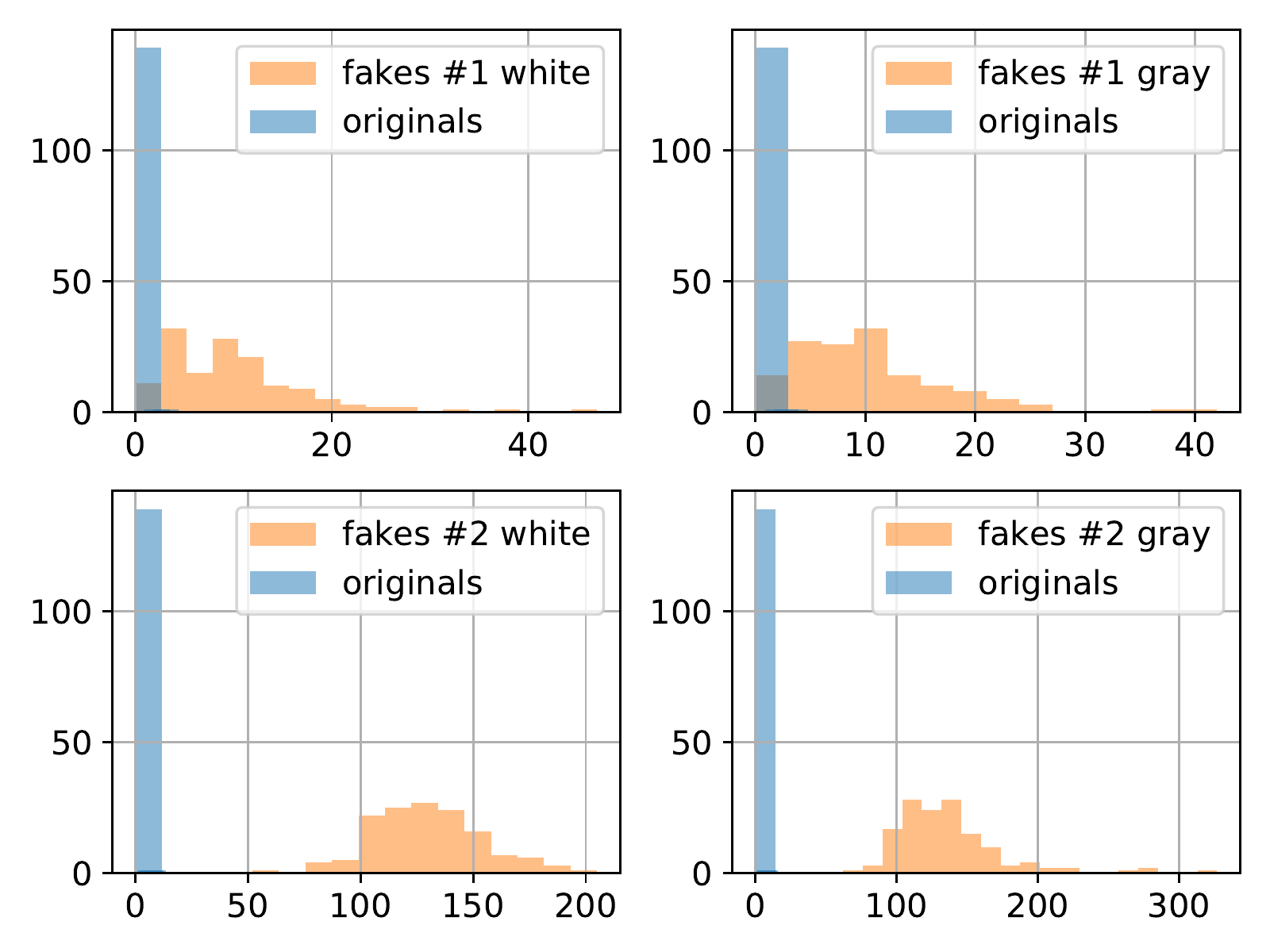}	
	\caption{The first scenario results' visualization: the histogram of symbol-wise Hamming distance (horizontal axis) between the original digital templates $\t$ and the corresponding estimations $\hat{\t}$ obtained via the encoder model trained with respect to the term $\Dtt$.}
	\label{fig:dtt haming dists distribution}
\end{figure}

\subsubsection{First scenario} 
\label{ch4_subsubsec:dtt}

The optimization problem based on $\loss_\textrm{One-class}^1(\bphi, \btheta) = -\Dtt$ aims at producing an accurate estimation $\hat{\t}$ of the corresponding binary digital template $\t$ for each input printed original code $\x$. Taking into account that due to the nature of the used trained model the output estimation is real valued but not binary, at the inference stage, to measure the Hamming distance the final estimation $\hat{\t}$ is obtained by the thresholding with a threshold 0.5. 

Fig. \ref{fig:dtt haming dists distribution} illustrates the distributions of the symbol-wise Hamming distance between the original digital templates $\t$ and the corresponding estimations $\hat{\t}$ obtained from the printed original and fake codes. Taking into account that the extracted feature vector consists only of one value, the OC-SVM is not used and the classification is performed based on the decision rule:
%
\begin{equation}
\left\{
\begin{array}{lll}
P_{fa}   & = & \textrm{Pr}\{d_{\textrm{\scriptsize{Hamming}}}(\t, \hat{\t}) \le \gamma_1    \;|\; \mathcal{H}_0 \}, \\
P_{miss} & = & \textrm{Pr}\{ d_{\textrm{\scriptsize{Hamming}}}(\t,\hat{\t})> \gamma_1 \;|\; \mathcal{H}_1\},
\end{array}
\right.
\label{ch4_eq:one metric decision}
\end{equation}
where $P_{miss}$ is a probability of miss and $P_{fa}$  is probability of false acceptance. The hypothesis $\mathcal{H}_0$ corresponds to the hypothesis that the input code is fake and the $\mathcal{H}_1$ corresponds to the hypothesis that the input code is original. Aiming to have $P_{miss} = 0$, the decision threshold $\gamma_1$ is determined on the validation sub-set to be equal to 2. The obtained classification error is given in Table \ref{tab:ib oc classification results}.

\begin{figure}[t!]
    \centering
	\includegraphics[width=0.5\linewidth]{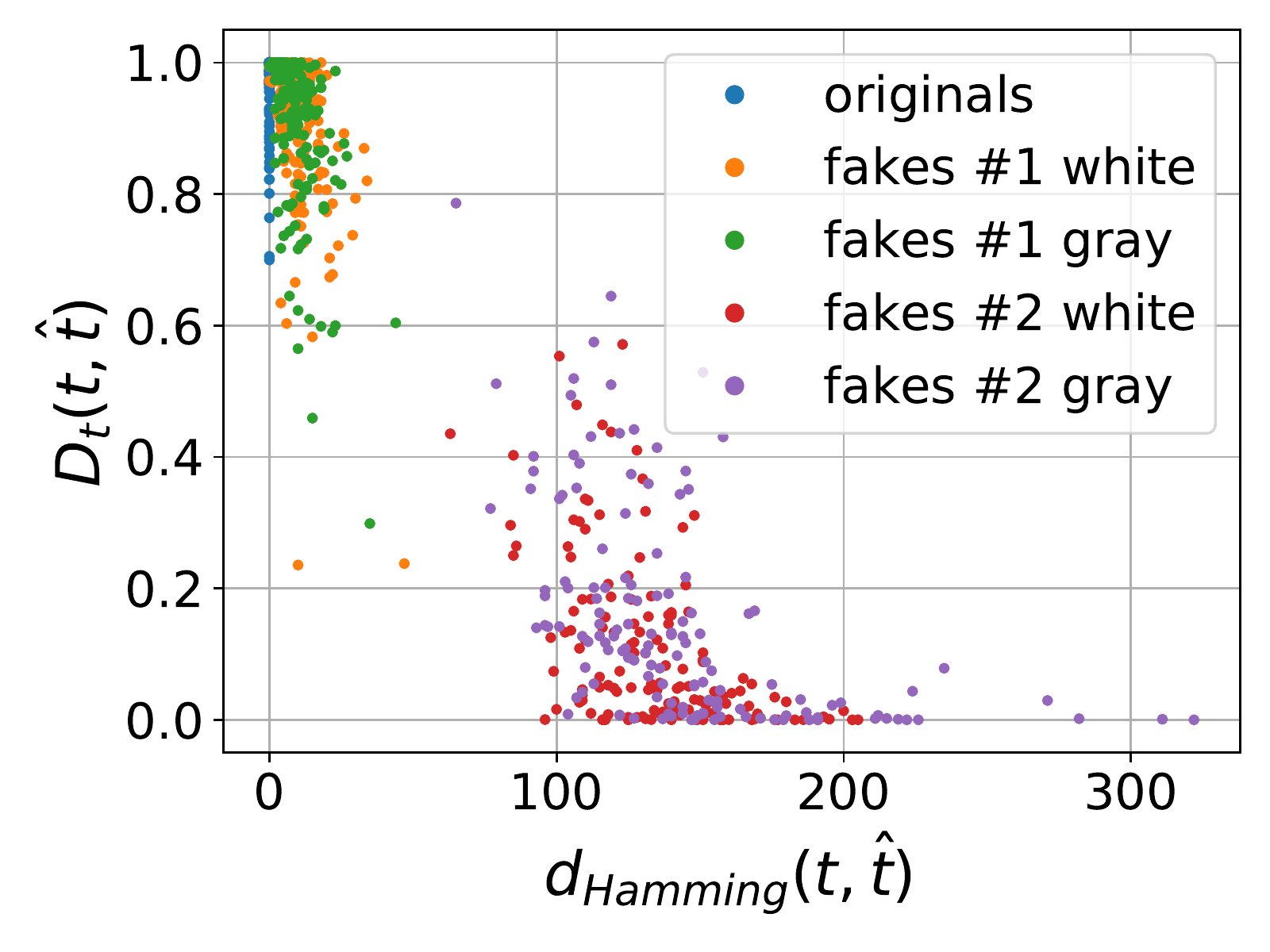}
	\caption{The second scenario results' visualisation: the 2D distribution of \textit{(i)} the symbol-wise Hamming distance between the original digital templates $\t$ and the corresponding estimations $\hat{\t}$ obtained via the encoder model trained with respect to the $\Dtt$ term and \textit{(ii)} the corresponding responses of the discriminator model trained with respect to the $\Dt$ term.}
	\label{fig:dtt dt}
\end{figure}

According to the obtained results, the one-class classification based on the encoder model trained with respect to the $\Dtt$ term as shown in Fig. \ref{fig:dtt dt dxx dx hc scheme} allows to distinguish the originals and the fakes \#2 with 100\% accuracy. The obtained $P_{miss}$ and $P_{fa}$ are confirmed by the distribution of the Hamming distance shown in Fig. \ref{fig:dtt haming dists distribution}. In case of the fakes \#1, the corresponding distributions overlap and the $P_{fa}$ is about 6 - 8\%.

\subsubsection{Second scenario}
\label{ch4_subsubsec:dtt dt}

The optimization problem based on $\loss_\textrm{One-class}^2(\bphi, \btheta) = -\Dtt + \Dt$ is an extension of the scenario \ref{ch4_subsubsec:dtt} with the discriminator part $\Dt$ that aims to distinguish between the distribution of original digital templates and its corresponding estimate. 

Fig. \ref{fig:dtt dt} presents the 2D distribution of \textit{(i)} the symbol-wise Hamming distance between the original digital templates $\t$ and the corresponding estimations $\hat{\t}$ obtained based on the encoder model trained with respect to the $\Dtt$ term and \textit{(ii)} the corresponding responses of the discriminator trained with respect to the $\Dt$ term as shown in Fig. \ref{fig:dtt dt dxx dx hc scheme}. It is easy to see that the obtained results are very close to those in Fig. \ref{fig:dtt haming dists distribution} with respect to the Hamming distance, namely, the results for the original codes are close to zero and overlap with the fakes \#1, while the fakes \#2 are well separable. With respect to the $\Dt$ discriminator decision the situation is similar, namely, the fakes \#2 are well separable by the decision ratio smaller then 0.5 - 0.6. At the same time, for the the fakes \#1 the decision ratio is bigger than 0.7 - 0.8 as well as for the originals. 

The obtained authentication error based on the $P_{miss}$ and $P_{fa}$ calculated with respect to the decision rule (\ref{ch4_eq:one metric decision}) and given in Table \ref{tab:ib oc classification results} shows that the regularization via the discriminator $\Dt$ does not have any significant influence and does not allow to improve the authentication accuracy. 

\subsubsection{Third scenario}
\label{ch4_subsubsec:dtt dxx}

\begin{figure}[t!]
    \centering
	\begin{subfigure}{0.48\columnwidth}    
		\includegraphics[width=1\linewidth,valign=t]{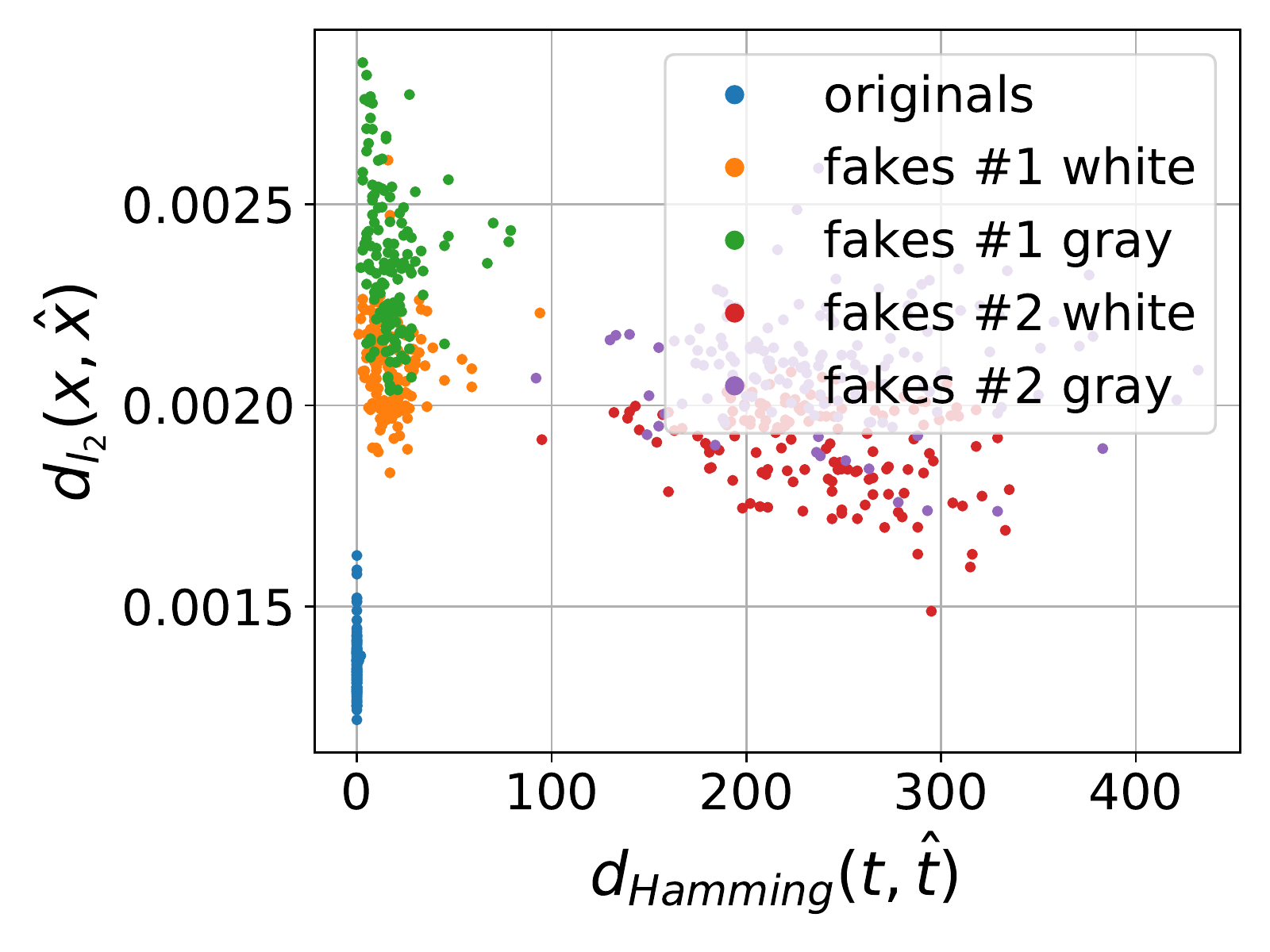}
		\captionsetup{justification=centering}
		\caption{}
		\label{fig:dtt dxx}
	\end{subfigure}
	\begin{subfigure}{0.48\columnwidth}    
		\includegraphics[width=1\linewidth,valign=t]{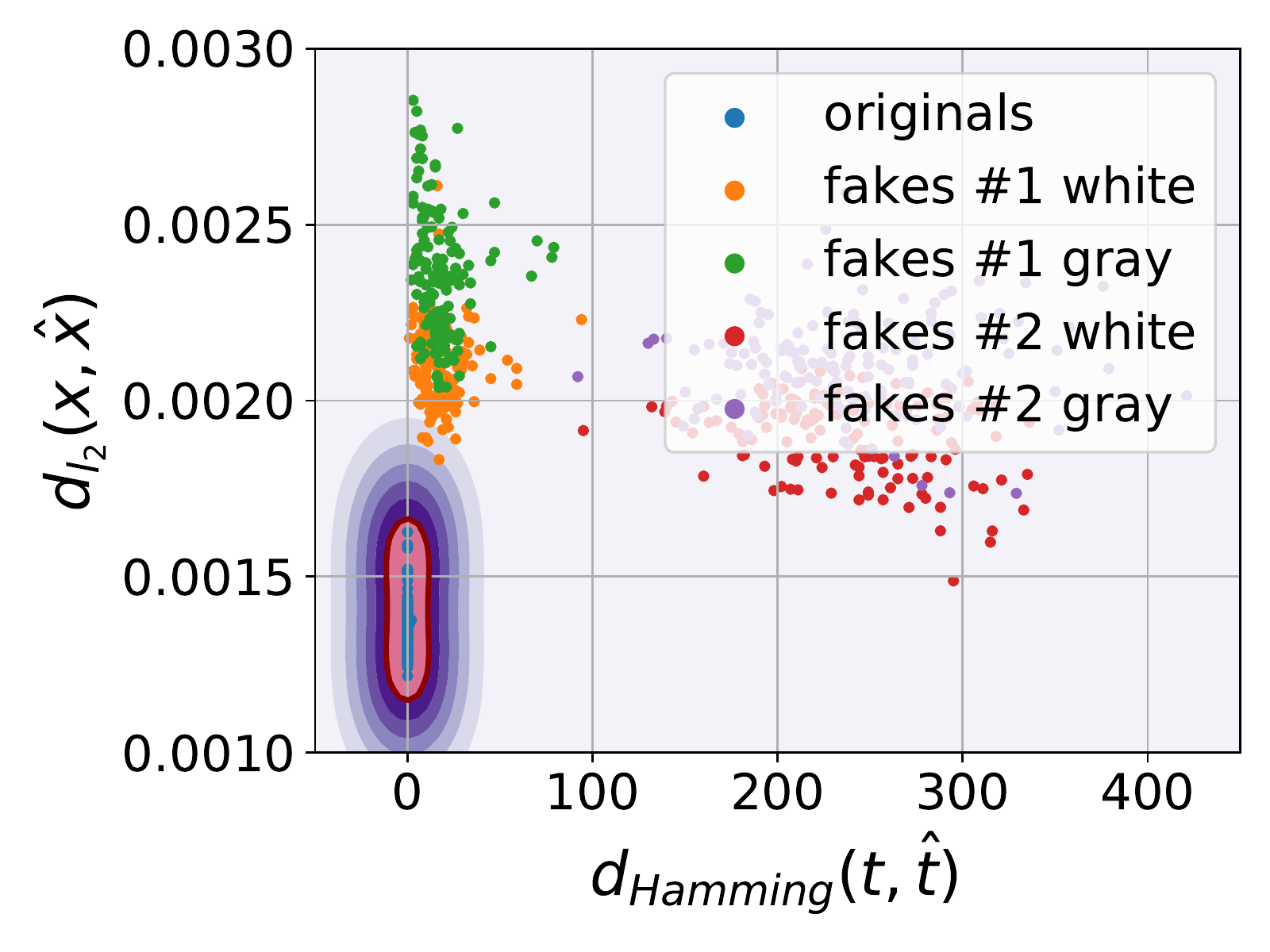}
		\captionsetup{justification=centering}
		\caption{}
		\label{fig:dtt dxx oc-svm}
	\end{subfigure}
	\caption{The third scenario results' visualization: (a) the distribution of \textit{(i)} the symbol-wise Hamming distance between the digital templates and its corresponding estimations via the encoder model trained with respect to the $\Dtt$ term and \textit{(ii)} the $\ell_2$ distance between the printed codes and its corresponding reconstructions by the decoder model trained with respect to the $\Dxx$ term; (b) the OC-SVM decision boundaries.}
	\label{fig:third scenario visualisation}
\end{figure}

In the third scenario $\loss_\textrm{One-class}^3(\bphi, \btheta) = -\Dtt - \beta \Dxx$ the term $\Dxx$ is in charge of the printed codes reconstruction and plays a role of a learnable regularization. 

Fig. \ref{fig:dtt dxx} demonstrates the obtained distribution of two metrics: \textit{(i)} the symbol-wise Hamming distance introduced in the Section \ref{ch4_subsubsec:dtt} and \textit{(ii)} the $\ell_2$ error between the printed codes and the corresponding reconstructions obtained as an output of the decoder model trained with respect to the $\Dxx$ term as shown in Fig. \ref{fig:dtt dt dxx dx hc scheme} without any additional post-processing. 

The obtained authentication results based on the decision rule (\ref{ch4_eq:one metric decision}) are given in Table \ref{tab:ib oc classification results}. It is easy to see that the learnable regularization via $\Dxx$ term preserves the $P_{miss}$ and $P_{fa}$ on the fakes \#2 to be zero, similar to the previous scenarios. At the same time, it allows to decrease the $P_{fa}$ for the fakes \#1 from 7\% till 1-1.6\%. Additionally, Table \ref{tab:ib oc classification results} presents the authentication results obtained based on the two metrics decision rule:
\begin{equation}
\left\{
\begin{array}{lll}
    P_{fa}   & = & \textrm{Pr}\{d_{\textrm{\scriptsize{Hamming}}}(\t, \hat{\t}) \le \gamma_1 \;\&\; \\ 
    & & \;\;\;\;\;\;\;\;\;\;\;\;  d_{\ell_2}(\x, \hat{\x}) \le \gamma_2   \;|\; \mathcal{H}_0 \} \\[0.25cm]
    P_{miss} & = & \textrm{Pr}\{ d_{\textrm{\scriptsize{Hamming}}}(\t, \hat{\t})> \gamma_1 \;\&\; \\ 
    & & \;\;\;\;\;\;\;\;\;\;\;\;  d_{\ell_2}(\x, \hat{\x}) > \gamma_2 \;|\; \mathcal{H}_1\}, 
\end{array}
\right.
\label{ch4_eq:two metric decision}
\end{equation}
that allows to significantly reduce the $P_{fa}$ for the fakes \#1 to about 0.28\%. Aiming to have the $P_{miss} = 0$, the decision constant $\gamma_2$ is determined on the validation sub-set to be equal 0.0017 and $\gamma_1$ equals to 2. 

In addition, Table \ref{tab:ib oc classification results} includes the results of OC-SVM trained with respect to the metrics under investigation (the symbol-wise Hamming distance between the digital templates and its corresponding estimations via the encoder model trained with respect to the $\Dtt$ term and the $\ell_2$ distance between the printed codes and its corresponding reconstructions by the decoder model trained with respect to the $\Dxx$ term). The OC-SVM is trained only on the train sub-set of the original printed codes $\x$ and its corresponding templates $\t$. The example of OC-SVM decision boundaries is illustrated in Fig. \ref{fig:dtt dxx oc-svm}. The OC-SVM reduces $P_{fa}$ to  0\% for all types of fakes. However, $P_{miss}$ increases to about 0.28\% in contrast to the previously obtained results with $P_{miss} = 0\%$. 

\begin{figure}[t!]
    \centering
	\begin{subfigure}{0.48\columnwidth}    
		\includegraphics[width=1\linewidth,valign=t]{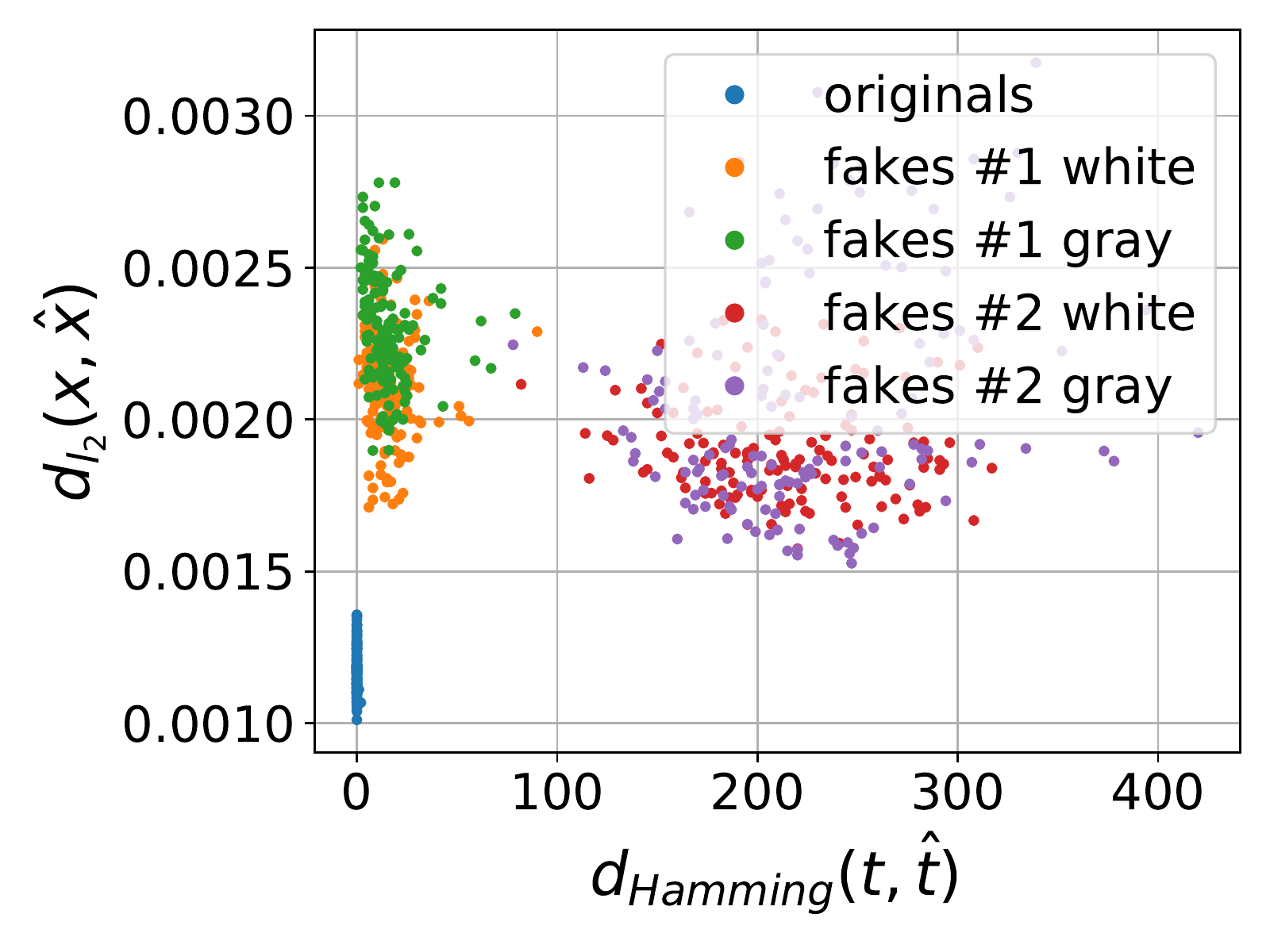}
		\captionsetup{justification=centering}
		\caption{}
		\label{fig:dtt dt dxx dx}
	\end{subfigure}
	\begin{subfigure}{0.48\columnwidth}    
		\includegraphics[width=1\linewidth,valign=t]{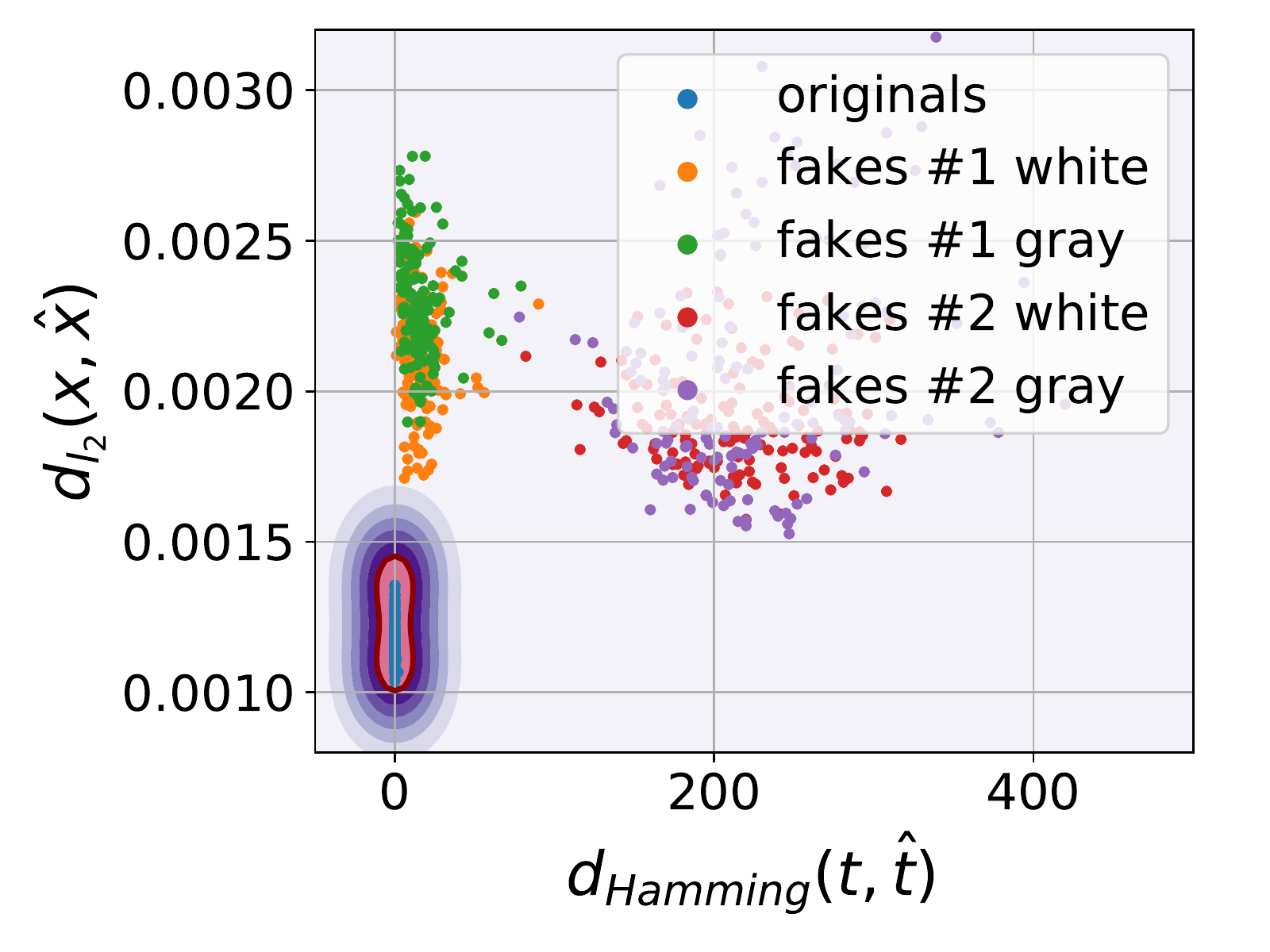}
		\captionsetup{justification=centering}
		\caption{}
		\label{fig:dtt dt dxx dx oc-svm}
	\end{subfigure}
	\caption{The fourth scenario results' visualization: (a) the distribution of \textit{(i)} the symbol-wise Hamming distance between the digital templates and its corresponding estimations via the encoder model trained with respect to the $\Dtt$ term and \textit{(ii)} the $\ell_2$ distance between the printed codes and its corresponding reconstructions by the decoder model trained with respect to the $\Dxx$ term; (b) the OC-SVM decision boundaries.}
	\label{fig:fourth scenario visualisation}
\end{figure}

\subsubsection{Fourth scenario}
\label{ch4_subsubsec:dtt dt dxx dx}

The last considered scenario $\loss_\textrm{One-class}^4(\bphi, \btheta) = -\Dtt + \Dt - \beta \Dxx + \beta \Dx$ includes four terms: the main term $\Dtt$, the discriminator $\Dt$ on the digital template estimation space, the printed code reconstruction space regularization $\Dxx$ and the discriminator $\Dx$. 
Similarly to the third scenario, the OC-SVM is trained with respect to the two features: \textit{(i)} the symbol-wise Hamming distance between the original digital templates  and their estimations and \textit{(ii)} the $\ell_2$ distance between the printed codes and their reconstructions. A visual representation of the jount distribution of these metrics is shown in Fig. \ref{fig:dtt dt dxx dx}. Table \ref{tab:ib oc classification results} includes the obtained one-class classification error based on three criteria: the decision rules (\ref{ch4_eq:one metric decision}) and (\ref{ch4_eq:two metric decision}) and the OC-SVM. The example of OC-SVM decision boundaries is illustrated in Fig. \ref{fig:dtt dt dxx dx oc-svm}. 


\begin{table}[t!]
	\centering
	\renewcommand*{\arraystretch}{1.35}
	\caption{Execution time (hours) per 100 epochs on one NVIDIA GPU with  a learning rate 1e-4 for the considered scenarios.}
	\label{tab:ib oc classification exec time}
	{\small
	\begin{tabular}{lc} \hline
		Model & Execution time, hours \\[0.1cm] \hline 
		$\loss_\textrm{One-class}^1: -\Dtt$ & 2.78 - 3.05 \\ 
		$\loss_\textrm{One-class}^2: -\Dtt + \Dt$ & 5.12 - 5.25 \\
		$\loss_\textrm{One-class}^3: -\Dtt - \beta \Dxx$ & 5.56 - 5.83 \\
		$\loss_\textrm{One-class}^4: -\Dtt + \Dt - \beta \Dxx + \beta \Dx$ & 11.11 - 11.39 \\ \hline
	\end{tabular}
	}
\end{table}    

From the obtained results, one can note that in terms of decision rule (\ref{ch4_eq:one metric decision}), the regularization via $\Dt$ and $\Dx$ discriminators is counter-productive and  makes the classification error bigger in comparison with the third scenario. In case of the decision rule (\ref{ch4_eq:two metric decision}), the regularization leads to a significant increase of $P_{miss}$. At the same time, the OC-SVM allows to decrease $P_{miss}$ in two times, from 0.28\% to 0.14\% preserving $P_{fa}$ equals to zero for all types of fakes. 

In summary, it should be pointed out that despite the great performance of the fourth scenario's model its complexity is times higher compared with the other considered scenarios. The execution time complexity in hours per 100 training epochs is given in Table \ref{tab:ib oc classification exec time} for each scenario.

\section{Conclusion}

In this work, we investigate the authentication aspects of modern CDP with respect to the typical hand-crafted copy fakes. To simulate the real-life conditions, we created the Indigo mobile dataset of CDP printed on the industrial printer and enrolled it via the mobile phone under regular light conditions. 

The performed analysis of the base-line multi-class supervised classification of CDP reveals two important observations:
\begin{itemize}
	\item In the general case, the model trained in a supervised way is capable to distinguish with a high accuracy the original CDP from the fakes produced on modern copy machines, which use built-in smart morphological processing enhancing image quality and reducing the dot gain for further reproduction. 
	\item The quality of the fakes used for the training plays a very important role. The superior quality fakes closer to the original codes are of preference for the training and allow the model to authenticate the inferior quality fakes, even when the model does not see them during the training. In contrast, if the classifier is trained on the inferior quality fakes, then it is not capable to authenticate the superior quality fakes.
\end{itemize}

The performed analysis of CDP authentication based on the one-class classification shows that:
\begin{itemize}
	\item In view of the great similarity between the original and fake codes the authentication in the spatial domain \textit{(i)} is difficult with respect to the finding of right metrics and \textit{(ii)} is not reliable enough due to the high overlapping between the classes.
	\item The authentication with respect to the digital templates is more efficient compared to the authentication with respect to the physical references.
	\item Despite the original black-and-white nature of the CDP the authentication based on codes taken by the mobile phone in color mode is more efficient compared to the grayscale mode. 
	\item The authentication with respect to the DNN estimation of the digital templates and printed codes reconstruction is more efficient than the direct authentication with respect to the digital and printed codes in spatial domain.
\end{itemize}

The main disadvantage of the DNN based models is its high training complexity compared to the direct authentication in spatial domain. At the same time, at the inference stage, the trained models are equivalent in terms of authentication complexity to the authentication in spatial domain.

Besides the impressive performance of the one-class classification on real samples and mobile phone verification, it should be pointed out that the above analysis is done with respect to the typical HC copy attacks. In view of the widespread use of the ML technologies, the question about the robustness to the ML attacks is an important problem that we aim at investigating in our future work. 




%


\ifCLASSOPTIONcaptionsoff
  \newpage
\fi



%
\bibliographystyle{IEEEtran}
\bibliography{references.bib}

\end{document}